\documentclass[english,10pt,aps,prd,a4paper,preprintnumbers,floatfix,nofootinbib,showpacs,superscriptaddress, notitlepage]{revtex4-1} 
 \pdfoutput=1
\usepackage[usenames,dvipsnames]{color}  
\usepackage{graphicx}
\usepackage{caption}
\usepackage{subcaption}
\usepackage{amsfonts}
\usepackage[export]{adjustbox} 
\usepackage{amsmath}
\usepackage{amssymb}
\usepackage{amsfonts}
\usepackage[colorlinks=true,citecolor=darkred,urlcolor=darkred, pdfborder={0 0 0}]{hyperref}
\usepackage[normalem]{ulem}
\usepackage{braket}
\usepackage{float}
\usepackage{multirow}

\makeatletter
\def\p@subsection{}
\makeatother

\definecolor{darkred}{rgb}{0.6,0,0}

\definecolor{linkcolor}{rgb}{0,0,0.5}

\def\gsim{\raise0.3ex\hbox{$\;>$\kern-0.75em\raise-1.1ex\hbox{$\sim\;$}}}
\def\lsim{\raise0.3ex\hbox{$\;<$\kern-0.75em\raise-1.1ex\hbox{$\sim\;$}}}

\def\beqn#1{\begin{equation}\label{#1}}
\def\eeqn{\end{equation}}

\def\beqa#1{\begin{eqnarray}\label{#1}}
\def\eeqa{\end{eqnarray}}

\def\0nbb {$0\nu\beta\beta$ }

\def\Z2{$\mathcal{Z_2}$}

\newcommand {\ignore}[1]{}

\newcommand{\sm}{{Standard Model }}

\def\321{$\mathrm{SU(3) \otimes SU(2) \otimes U(1)}$ }

\newcommand{\AddrHBNI}{
	Homi Bhabha National Institute, BARC Training School Complex, Anushakti Nagar, Mumbai 400094, India }

\begin{document}

\bibliographystyle{unsrt} 

\title{Fermionic Dark Matter in Dynamical Scotogenic Model}
\author{Eung Jin Chun}
\email{ejchun@kias.re.kr}
\affiliation{Korea Institute for Advanced Study, Seoul 02455, Korea}
\author{Abhishek Roy}
\email{abhishek.r@iopb.res.in}
\affiliation{Institute of Physics, Sachivalaya Marg, Bhubaneswar, Pin-751005, Odisha, India}
\affiliation{\AddrHBNI}
\author{Sanjoy Mandal}
\email{smandal@kias.re.kr}
\affiliation{Korea Institute for Advanced Study, Seoul 02455, Korea}
\author{Manimala Mitra}
\email{manimala@iopb.res.in}
\affiliation{Institute of Physics, Sachivalaya Marg, Bhubaneswar, Pin-751005, Odisha, India}
\affiliation{\AddrHBNI}

\preprint{IP/BBSR/2023-03}

\begin{abstract}
In the Dynamical Scotogenic Model, the global $B-L$ symmetry is supposed to be broken spontaneously resulting in a massless Goldstone boson called majoron, and massive right handed neutrinos which participate in the generation of light neutrino massses at one-loop. One of them being the lightest stable particle can be a thermal dark matter candidate.  We discuss how  the dark matter phenomenology differs from the original Scotogenic model, taking  into account all the constraints coming from the observed neutrino masses and mixing,  lepton flavor violations such as $\mu \to e\gamma, \mu \to e J$,  astrophysical and cosmological observations of stellar cooling and $N_{eff}$, as well as collider signatures such as Higgs invisible decays. We find  that the dark matter annihilation to majorons plays an important role to produce the right relic abundance.
\end{abstract}
\maketitle
\section{Introduction}
One of the strongest arguments in favor of new physics is the existence of light Standard Model (SM) neutrinos with non-zero masses and mixings~\cite{Kajita:2016cak,McDonald:2016ixn,KamLAND:2002uet,K2K:2002icj}. A plethora of neutrino oscillation experiments have given conclusive evidence that  the mass square differences of the light neutrinos are $\Delta m^2_{12} \sim 10^{-5}$ and $|\Delta m^2_{13}| \sim 10^{-3}$ and the mixing angles are 
$\theta_{12} \sim 34.3^{\circ}, \theta_{23} \sim 49.26^{\circ}$ and  $\theta_{13} \sim 8.53^{\circ}$. The origin of neutrino masses and mixings stands out as one of the biggest challenges of modern elementary particle physics, which requires satisfactory explanation. The existence of dark matter (DM) is another open problem of particle physics, the basic understanding and interpretation of which is still  lacking ~\cite{Bertone:2004pz}.  A number of different experimental observations including observation of the flatness of the rotation curve of galaxies, gravitational lensing, and the Cosmic Microwave Background (CMB) radiation by Planck have given indication that the present DM relic density is $\Omega h^2_{DM}=0.1199 \pm 0.0027$~\cite{Planck:2015fie}.  One of the most appealing neutrino mass  generation paradigms is the Scotogenic extension ~\cite{Ma:2006km} of the SM, where  neutrino mass generation is inherently  connected with DM. 
The  model contains a scalar doublet $\eta$ and three copies of SM singlet fermions, $N_i$, both are odd under a  $\mathbb{Z}_2$ symmetry. Neutrino mass is  generated at one-loop with the $\mathbb{Z}_2$-odd particles mediating  in the loop. The same $\mathbb{Z}_2$ symmetry is also responsible for the stability of the DM. Depending on the mass hierarchy between $\eta$ and $N$,  the DM in this model  can  either  be fermionic~ $N$ or bosonic $\eta$. This strong connection between DM and neutrino mass generation has led to many interesting studies~\cite{FileviezPerez:2019cyn,Escribano:2021ymx,Ma:2021eko,Babu:2007sm,Mandal:2019oth,DeRomeri:2022cem,Portillo-Sanchez:2023kbz,Escribano:2021ymx,DeRomeri:2021yjo,Escribano:2020iqq,Escribano:2020wua}. The other appealing variation of the Scotogenic model is referred as dynamical Scotogenic extension \cite{Bonilla:2019ipe,DeRomeri:2022cem}, which in addition to  $\eta$ and $N$'s also contains another $SU(3)_c\otimes SU(2)_L\otimes U(1)_Y$ singlet scalar $\sigma$. In this model a global $U(1)$ symmetry is spontaneously broken down  via the vacuum expectation value~(VEV) of $\sigma$, which generate the Majorana masses of the $N$'s. 
\par The purpose of this paper is to examine the prospects of fermionic DM for  dynamical Scotogenic model~\cite{Bonilla:2019ipe,DeRomeri:2022cem}. In the basic Scotogenic model (referred in this paper as 'vanilla Scotogenic'), Majorana mass term $M_N\overline{N^c}N$ is explicit, whereas in the dynamical version of Scotogenic model, Majorana mass term is generated via the vacuum expectation value~(VEV) of a $SU(3)_c\otimes SU(2)_L\otimes U(1)_Y$ singlet scalar $\sigma$. This dynamical version harbours a physical Nambu-Goldstone boson~(NGB), dubbed as majoron $J$~\cite{Chikashige:1980ui,Schechter:1981cv,Konoplich:1988mj} and one extra CP-even scalar $H$. The existence of majoron induces  a number of  distinct observables such as Higgs invisible decay to a pair of majoron, as well as, charged lepton flavour violating decays $\ell_\beta\to\ell_\alpha\gamma$, $\ell_\beta\to\ell_\alpha J$.
\par In the original version of Scotogenic model, the only DM annihilation is $t$-channel process $N_i N_i\to\overline{\ell_\beta}\ell_\alpha/\nu_\beta\nu_\alpha$ which are mediated by $\mathbb{Z}_2$-odd particles. This involves the same Yukawa coupling which is also responsible for charged lepton flavour violations~(cLFV). As the Yukawa coupling is strongly constrained by bounds on cLFV processes, the DM will be overabundant in large region of parameter space  unless  the co-annihilation effects are taken into account. The  dynamical Scotogenic model, because of the presence of additional DM  annihilation channels such as $N_iN_i\to JJ, Jh/H$ and $N_i N_i\to \text{SM} \text{ SM}$, where SM stands for SM particles,
facilitates to overcome this problem.  We find that these new channels help to satisfy the correct relic abundance for DM mass below few TeV, in large parameter space.
In DM analysis, we take into account bounds from the cLFV, various SM Higgs measurements, DM direct detection as well as astrophysical measurements coming from stellar cooling and the effective number of neutrino species before recombination.
\par The paper is organized as follows: we first briefly present the model in the next section, where we discuss the particle content, scalar sector and neutrino mass generation at one-loop. In Sec.~\ref{sec:collider}, we discuss the existing collider constraints coming from measurements of Higgs invisible decay, signal strength parameter at LHC, $W$ and $Z$ boson decay width measurements at LEP-I and charged Higgs searches at LEP-II. In Sec.~\ref{sec:cLFVandStellar}, we discuss the constraints coming from cLFV and astrophysical observables such as stellar cooling and $N_{eff}$, while the results of our DM analysis are presented in Sec.~\ref{sec:DManalysis}. Finally, in Sec.~\ref{sec:conclusion}, we conclude.
\section{Dynamical Scotogenic Model}
\label{sec:model}
We  briefly discuss the basic features of the dynamical Scotogenic model which was previously discussed in Ref.~\cite{DeRomeri:2022cem,Bonilla:2019ipe}. In addition to SM particle content the model contains  three SM singlet fermions $N_i$, one inert scalar doublet $\eta$ with hypercharge $1/2$ and a complex scalar $\sigma$. The new particles and their quantum numbers under SM gauge group $SU(3)_c\otimes SU(2)_L\otimes U(1)_Y$ are given in Table.~\ref{tab:particle-content}, where the index $i$ runs from 1 to 3. The gauge singlet fermionic state $N_1$ is the lightest stable particle in our consideration and hence a suitable choice for DM. The additional $\mathbb{Z}_2$ symmetry is the ``dark parity" responsible for the stability of the DM candidate. 
\begin{table}[h]
	\centering
	\begin{tabular}{|c|c|c|c|c|c|c|}
		\hline \phantom{XXXXXXXX} & \phantom{X} $L_i$ \phantom{X} & \phantom{X}$\ell_{R_i}$\phantom{X} & \phantom{X} $\Phi$ \phantom{X} & \phantom{X} $\eta$ \phantom{X} & \phantom{X} $N_i$ \phantom{X} & \phantom{X} $\sigma$ \phantom{X} \\
		\hline \hline
		{\bf  $SU(2)_L$} & {\bf  $2$} & {\bf  $1$} & {\bf  $2$} & {\bf  $2$} & {\bf  $1$} & {\bf  $1$} \\
		\hline
		{\bf  $U(1)_Y$} & {\bf  $-1/2$} & {\bf  $-1$} & {\bf  $1/2$} & {\bf  $1/2$} & {\bf  $0$} & {\bf  $0$} \\
		\hline
		{ $U(1)_{B-L}$} & {$-1$} & {$-1$} & {$0$} & {$0$} & {$-1$} & {$2$}  \\
		\hline
		$\mathbb{Z}_2$ & $+$ & $+$ & $+$  & $-$ & $-$ & $+$ \\
		\hline
	\end{tabular}
	\caption{\centering Particle content and charge assignments of the dynamical Scotogenic model. \label{tab:Dynamical-Scotogenic-Model}}
	\label{tab:particle-content}
\end{table}
Under this dark $\mathbb{Z}_2$ parity, all the SM particles and complex scalar $\sigma$ are even, whereas the dark sector, which consists of fermions $N_i$ and doublet $\eta$, is odd. In addition to these symmetries, the model also contains a global $U(1)_{B-L}$ symmetry, which is being spontaneously broken. The scalar singlet $\sigma$ has a non-trivial charge under global $U(1)_{B-L}$, and its VEV $v_{\sigma}$ breaks global $U(1)_{B-L}$. The most general renormalizable and $SU(3)_c\otimes SU(2)_L\otimes U(1)_Y\otimes U(1)_{B-L}\otimes \mathbb{Z}_2$ gauge invariant Yukawa Lagrangian can be written as
\begin{equation}
-\mathcal{L}_{\text{Y}} \supset Y^{\ell}_{ij} \bar{L}_i \Phi \ell_{R_j} + Y^{\nu}_{ij} \, \bar{L}_i  \tilde{\eta} N_j + \cfrac{1}{2} Y^N_{ij} \, \sigma \bar{N}^c_i N_j + \textit{h.c.},
\label{eq:LagYL}
\end{equation}
where $\tilde{\eta}=i\sigma_2\eta^*$ and $L=(\nu_{\ell L},\ell_L)^T$. The relevant $SU(3)_c\otimes SU(2)_L\otimes U(1)_Y\otimes U(1)_{B-L}\otimes \mathbb{Z}_2$ gauge invariant scalar potential that can break both the electroweak gauge symmetry as well as lepton number is given as
\begin{align}
V &=  m_\Phi^2 \Phi^\dagger \Phi + m_\eta^2 \eta^\dagger \eta + m_\sigma^2 \sigma^* \sigma + \lambda_\Phi (\Phi^\dagger \Phi)^2 + \lambda_\eta (\eta^\dagger \eta)^2 + \lambda_3 (\eta^\dagger \eta)(\Phi^\dagger \Phi) + \lambda_4 (\eta^\dagger \Phi) (\Phi^\dagger \eta ) \nonumber \\
&+ \displaystyle \frac{\lambda_5}{2} \left[ (\eta^\dagger \Phi)^2 + \text{h.c} \right] + \lambda_\sigma (\sigma^* \sigma)^2 + \lambda_{\Phi\sigma} (\Phi^\dagger \Phi)(\sigma^* \sigma) + \lambda_{\eta\sigma} (\eta^\dagger \eta) (\sigma^* \sigma).
\label{eq:potential-dynamical}
\end{align}
For definiteness, in the above, all parameters are assumed to be real. To guarantee that the scalar potential is bounded from below and has a stable vacuum at any given energy scale, the following conditions must hold~\cite{Kadastik:2009cu}
\begin{align}
& \lambda_{\Phi}, \lambda_{\eta}, \lambda_\sigma > 0, \,\, \lambda_3 > -2\sqrt{\lambda_\Phi \lambda_\eta},\,\, \lambda_3 + \lambda_4 - |\lambda_5| > -2 \sqrt{\lambda_\Phi\lambda_\eta},\nonumber \\
& 4\lambda_\Phi\lambda_\sigma > \lambda_{\Phi\sigma}^2,\,\, 4\lambda_\eta\lambda_\sigma > \lambda_{\eta\sigma}^2.
\label{eq:stability}
\end{align}
Further we restrict the scalar quartic couplings in Eq.~\ref{eq:potential-dynamical} as $\lambda_{i}< 4\pi$ in order to guarantee perturbativity.
 \par In order to ensure DM stability the $\mathbb{Z}_2$ symmetry should remain unbroken. This implies that the $\mathbb{Z}_2$ odd scalar doublet $\eta$ should not acquire a nonzero VEV. As a result, electroweak symmetry breaking and $B-L$ symmetry breaking is driven simply by the VEV of $\Phi$ and $\sigma$, respectively. In order to obtain the mass spectrum for the scalars after gauge and $B-L$ symmetry breaking, we expand the scalar fields as:
\begin{align}
\Phi=
\begin{pmatrix}
\phi^+\\
(v_\Phi+h_\Phi+i\eta_\Phi)/\sqrt{2}
\end{pmatrix},\,\,
\eta=
\begin{pmatrix}
\eta^+\\
(\eta^R+i\eta^I)/\sqrt{2}
\end{pmatrix}\,\,\text{and}\,\, \sigma=(v_\sigma+h_\sigma+i J)/\sqrt{2}.
\label{eqmass}
\end{align}
Note that the mixing of the Higgs $\Phi$ and the dark doublet $\eta$ is prohibited by the exact conservation of the $\mathbb{Z}_2$ symmetry. The components of $\eta$ have the following masses
\begin{align}
 m_{\eta^{R}}^{2}&=m_{\eta}^{2}+\frac{1}{2}\lambda_{\eta\sigma}v_\sigma^2+\frac{1}{2}\left(\lambda_{3}+\lambda_{4}+\lambda_{5}\right)v_\Phi^{2},\\
  m_{\eta^{I}}^{2}&=m_{\eta}^{2}+\frac{1}{2}\lambda_{\eta\sigma}v_\sigma^2+\frac{1}{2}\left(\lambda_{3}+\lambda_{4}-\lambda_{5}\right)v_\Phi^{2},\\
 m_{\eta^{+}}^{2}&=m_{\eta}^{2}+\frac{1}{2}\lambda_{\eta\sigma}v_\sigma^2+\frac{1}{2}\lambda_{3}v_\Phi^{2}.
\end{align}
The difference $m_{\eta^R}^2-m_{\eta^I}^2$ depends only on the parameter $\lambda_5$ which, we will show later, is also responsible for smallness of neutrino mass.
\par The scalar sector that results from Eq.~\ref{eq:potential-dynamical} gives rise to the majoron $J=\text{Im }(\sigma)$, a physical massless Goldstone boson, as well as two massive neutral CP-even scalars, $h$, and $H$. The mass matrix of  CP-even Higgs  scalars in the basis $(h_\Phi,h_\sigma)$ reads as~\cite{Joshipura:1992hp} 
\begin{align}
M_R^2=
\begin{bmatrix}
2\lambda_\Phi v_\Phi^2  &  \lambda_{\Phi\sigma}v_\Phi v_\sigma \\
\lambda_{\Phi\sigma} v_\Phi v_\sigma & 2\lambda_\sigma v_\sigma^2 
\end{bmatrix},
\end{align}
with the mass eigenvalues given by 
\begin{align}
m_{h,H}^2 &=\lambda_\Phi v_\Phi^2 +\lambda_\sigma v_\sigma^2 \mp \sqrt{(\lambda_\Phi v_\Phi^2 - \lambda_\sigma v_\sigma^2)^2+\lambda_{\Phi \sigma}^2v_\Phi^2 v_\sigma^2},
\end{align}
where by convention we choose $m_{h}^2<m_{H}^2$ throughout this work. $h$ is identified as the SM Higgs discovered at the  LHC. The two mass eigenstates $h,H$ are related with the $h_\Phi, h_\sigma$ fields through the rotation matrix $O_R$ as,  
\begin{align}
\begin{bmatrix}
h\\
H\\
\end{bmatrix}
=O_R
\begin{bmatrix}
h_\Phi\\
h_\sigma\\
\end{bmatrix}
=
\begin{bmatrix}
\cos\theta & \sin\theta \\
-\sin\theta & \cos\theta \\
\end{bmatrix}
\begin{bmatrix}
h_\Phi\\
h_\sigma\\
\end{bmatrix}\,\, \text{with}\,\, \tan 2\theta=\frac{\lambda_{\Phi\sigma}v_\Phi v_\sigma}{\lambda_\Phi v_\Phi^2-\lambda_\sigma v_\sigma^2}.
\label{eq:mixing-relation}
\end{align}
Additionally, we can also write the quartic couplings, $\lambda_\Phi$, $\lambda_\sigma$ and $\lambda_{\Phi\sigma}$ in terms of mixing angle $\theta$ and the scalar masses $m_{h,H}$ as follows:
\begin{align}
\lambda_\Phi =\frac{m_{h}^2\cos^2\theta+m_{H}^2\sin^2\theta}{2v_\Phi^2},
\lambda_\sigma =\frac{m_{h}^2\sin^2\theta+m_{H}^2\cos^2\theta}{2v_\sigma^2}\,\, \text{and}\,\,
\lambda_{\Phi\sigma} =\frac{\sin 2\theta (m_{h}^2-m_{H}^2)}{2 v_\Phi v_\sigma}.
\label{eq:lambdas}
\end{align}
One can see from Eq.~\ref{eq:lambdas}, how one can determine the quartic couplings from the knowledge of $m_H$, $v_\sigma$ and mixing angle $\theta$. The state $J$ in Eq.~\ref{eqmass} is a massless majoron state. As we will show later, this state plays major role in determining DM relic abundance, as well as, can induce lepton flavor violation $\mu \to e J$. We also discuss later how this presence of majoron can alter the stellar cooling mechanism and $N_{\rm eff}$.
\par
\begin{figure}[htb!]
\centering
\includegraphics[width=0.35\textwidth]{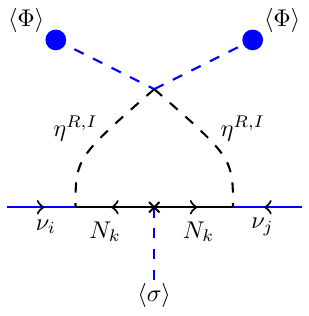}
\caption{\centering One loop Feynman diagram for neutrino mass generation.}
\label{fig:loop-neutrino}
\end{figure}

 Although the usual tree-level seesaw contribution to neutrino masses is forbidden by the $\mathbb{Z}_2$ symmetry, these are induced at the 1-loop through the exchange of the ``dark" fermions and scalar as illustrated in Fig.~\ref{fig:loop-neutrino}. This loop is calculable and neutrino mass is given by the following expression \cite{Mandal:2021yph,Ma:2006km,DeRomeri:2022cem,Vicente:2014wga}
\begin{align}
\label{eq:numass}
(m_\nu)_{ij} &
= \sum_{k=1}^3 \frac{Y_\nu^{ik} \, Y_\nu^{kj} M_{N_k}}{32 \pi^2}\left[ \frac{m_{\eta_R}^2}{m_{\eta_R}^2-M_{N_k}^2} \log \frac{m_{\eta_R}^2}{M_{N_k}^2} -  \frac{m_{\eta_I}^2}{m_{\eta_I}^2 - M_{N_k}^2} \log \frac{m_{\eta_I}^2}{M_{N_k}^2} \right],\\
& \equiv (Y_\nu^T \Lambda Y_{\nu})_{ij},
\end{align}
where $\Lambda$ matrix is defined as $\Lambda=\text{diag}(\Lambda_1,\Lambda_2,\Lambda_3)$, with
\begin{align}
\label{eq:Lambda}
\Lambda_k
=  \frac{M_{N_k}}{32 \pi^2}\left[ \frac{m_{\eta_R}^2}{m_{\eta_R}^2-M_{N_k}^2} \log \frac{m_{\eta_R}^2}{M_{N_k}^2} -  \frac{m_{\eta_I}^2}{m_{\eta_I}^2 - M_{N_k}^2} \log \frac{m_{\eta_I}^2}{M_{N_k}^2} \right].
\end{align}
Note that, $M_{N_k}=Y_N v_\sigma/\sqrt{2}$ is the Majorana masses of the $N_k$ states which are generated after the global $U(1)_{B-L}$ symmetry breaking. We choose to work in a basis where $M_N$ is diagonal, which implies a diagonal $Y_N$. Note that $m_{\eta^R}^2=m_{\eta^I}^2$ in the limit $\lambda_5\to 0$, which leads to an exact cancellation between the $\eta^R$ and $\eta^I$ loops, and vanishing neutrino masses. In the limit $\lambda_5\ll 1$, simplified expressions can be obtained~\cite{Vicente:2014wga}
\begin{align}
\label{eq:numass2}
(m_\nu)_{ij} 
\approx \frac{\lambda_5 v_\Phi^2}{32\pi^2}\sum_{k=1}^3 \frac{Y_\nu^{ik} \, Y_\nu^{kj} }{M_{N_k}}\left[ \frac{M_{N_k}^2}{m_{0}^2-M_{N_k}^2}  +  \frac{M_{N_k}^4}{(m_{0}^2 - M_{N_k}^2)^2} \log \frac{M_{N_k}^2}{m_0^2} \right],
\end{align}
where $m_{\eta^R}^2\approx m_{\eta^I}^2\equiv m_{0}^2$. From Eq.~\ref{eq:numass2} it is clear that one can fit the observed atmospheric and solar mass square differences in many ways as long as one takes an adequately small value for $\lambda_5$. For example, for sufficiently small $\lambda_5$ values, one can choose a reasonable Yukawa coupling $Y_\nu$ even for TeV scale dark sector particles $N_k$ and $\eta$. It's convenient to parameterize the Yukawa coupling using the Casas-Ibarra form~\cite{Casas:2001sr},
\begin{align}
Y_\nu=\sqrt{\Lambda}^{-1}R \sqrt{\hat{m}_\nu} U_{\rm lep}^{\dagger},
\label{eq:Ynu}
\end{align}
Here $R$ is a $3 \times 3$ complex orthogonal matrix, $U_{\text{lep}}$ is the leptonic mixing matrix which diagonalized the neutrino mass matrix as 
\begin{equation}
U_{\text{lep}}^{T} \, m_\nu \, U_{\text{lep}}=\hat{m}_\nu\equiv
\left(
\begin{array}{ccc}
m_1 & 0 & 0\\
0 & m_2 & 0\\
0 & 0 & m_3
\end{array}
\right) \, ,\quad \text{\bf for NO}
\label{eq:mnudiagNO}
\end{equation}
\begin{equation}
U_{\text{lep}}^{T} \, m_\nu \, U_{\text{lep}}=\hat{m}_\nu\equiv
\left(
\begin{array}{ccc}
m_3 & 0 & 0\\
0 & m_2 & 0\\
0 & 0 & m_1
\end{array}
\right) \, ,\quad \text{\bf for IO}
\label{eq:mnudiagIO}
\end{equation}
where $m_i$'s are the light neutrino masses and {\bf NO}~({\bf IO}) stand for normal~(inverted) ordering. The explicit form of the complex matrix $R$ is given as follows,
\begin{align}
R = \left( \begin{array}{ccc}  1 & 0 & 0
  \\ 0 & \cos x & \sin x \\
0 & -\sin x & \cos x\\
\end{array} \right)  \left( \begin{array}{ccc}  \cos y & 0 & \sin y
  \\ 0 & 1 & 0 \\
-\sin y & 0 & \cos y\\
\end{array} \right)  \left( \begin{array}{ccc}  \cos z & \sin z & 0
  \\ -\sin z & \cos z & 0 \\
0 & 0 & 1\\
\end{array} \right),
\end{align}
 where $x=\theta_{x}^{R}+ i\theta_{x}^{I}\ $,$y=\theta_{y}^{R}+ i\theta_{y}^{I}\ $ and $z=\theta_{z}^{R}+ i\theta_{z}^{I}$ are the complex angles with $\theta_{x,y,z}^{R},\theta_{x,y,z}^{I} \in R$. In our subsequent analysis, we follow normal mass hierarchy. For definiteness, we fix the neutrino oscillation parameters to their best-fit values \cite{deSalas:2020pgw}.
\section{Collider Constraints}
\label{sec:collider}
In this section we discuss the existing constraints on the relevant parameter space from various collider observables such as invisible Higgs decay, signal strength parameters and gauge boson decay widths. First, note that the 
coupling of the SM Higgs boson  to SM particles changes according to the following substitution rule due to the presence of the heavy Higgs $H$
\begin{align}
h_{\Phi}\to \cos\theta h- \sin\theta H 
\label{hsmcopling}
\end{align}
Morever when the scale associated with $U(1)_{B-L}$ violation is relatively low, the standard model Higgs boson $h$ can have potentially large invisible decays to majoron~\cite{Joshipura:1992hp}
\begin{align}
\Gamma(h\to JJ)&=\frac{1}{32\pi m_h}\frac{m_h^4\sin^2\theta}{v_\sigma^2}\sqrt{1-\frac{4m_{J}^2}{m_h^2}}.
\label{eq:inv-majoron}
\end{align}
In addition to the majoron channel, if the mass of fermionic DM $N_1$ is smaller than half of the Higgs mass, then the following channel will also contribute to the invisible decay
\begin{align}
\Gamma(h\to N_1 N_1)=\frac{\sin^2\theta }{16\pi m_h}\frac{M_{N_1}^2}{v_\sigma^2}\left(m_h^2-2M_{N_1}^2\right)\sqrt{1-\frac{4M_{N_1}^2}{m_h^2}}
\end{align}
Hence the total invisible decay width of \sm Higgs boson $h$ is given as 
\begin{align}
\Gamma^{\text{inv}}(h)=\Gamma(h\to JJ)+\theta(m_h-2M_{N_1})\Gamma(h\to N_1 N_1).
\label{eq:inv-Higgs}
\end{align}
Accordingly, the invisible branching ratio for $h$ is given by
\begin{align}
\text{BR}^{\text{inv}}(h)=
\frac{\Gamma^{\text{inv}}(h)}{\cos^2\theta\Gamma^\text{SM}(h)+\Gamma^{\text{inv}}(h)}.
\label{eq:inv-BR_Higgs}
\end{align}
The current upper limit on the branching ratio to invisible decay modes by the CMS experiment~\cite{CMS:2022qva} is:
\begin{align}
\text{BR}^{\text{inv}}(h) < 0.18\text{ at } 95\% \text{ C.L.}
\end{align}
\begin{figure}[htb!]
	\centering
	\includegraphics[width=0.45\textwidth]{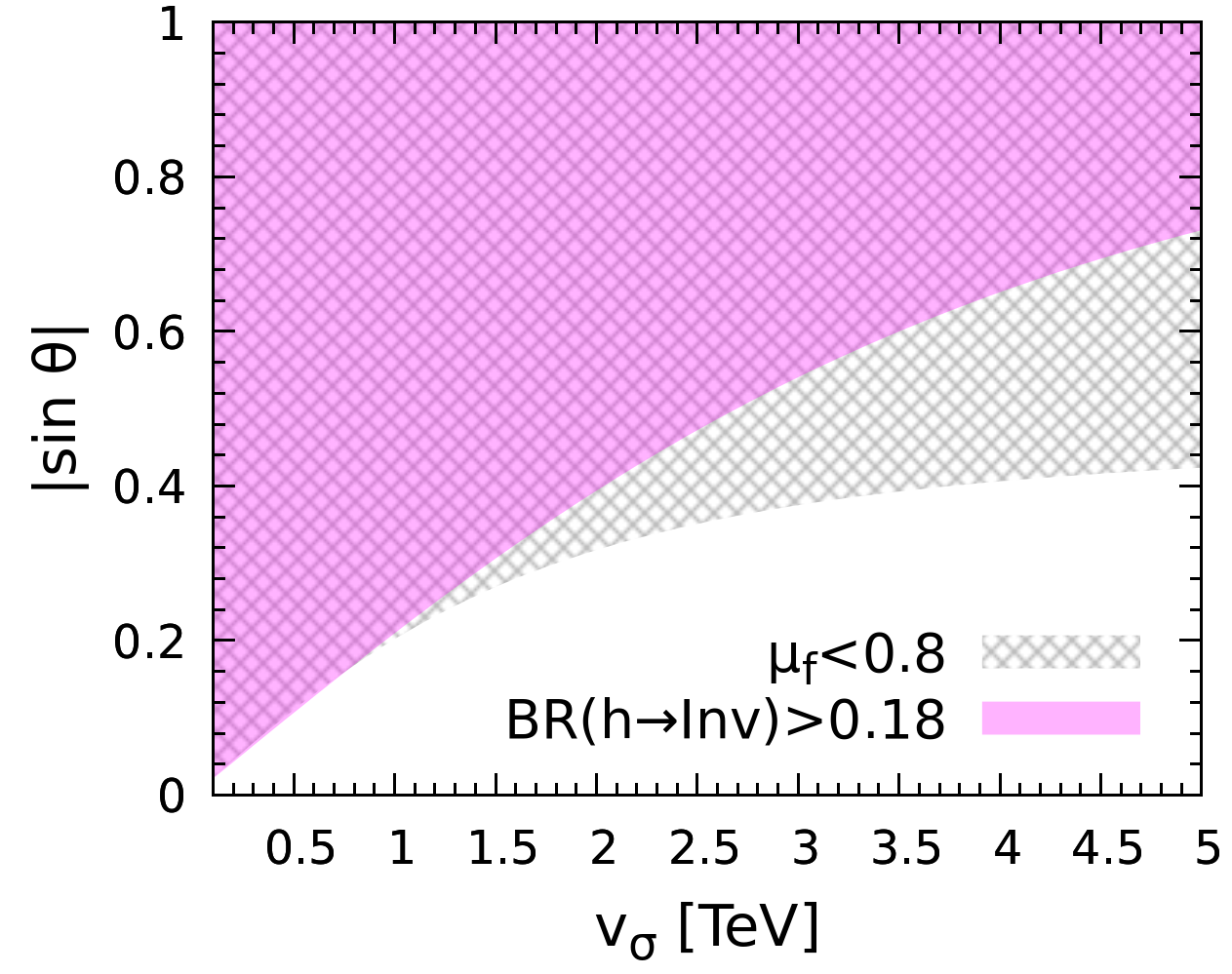}
	\includegraphics[width=0.45\textwidth]{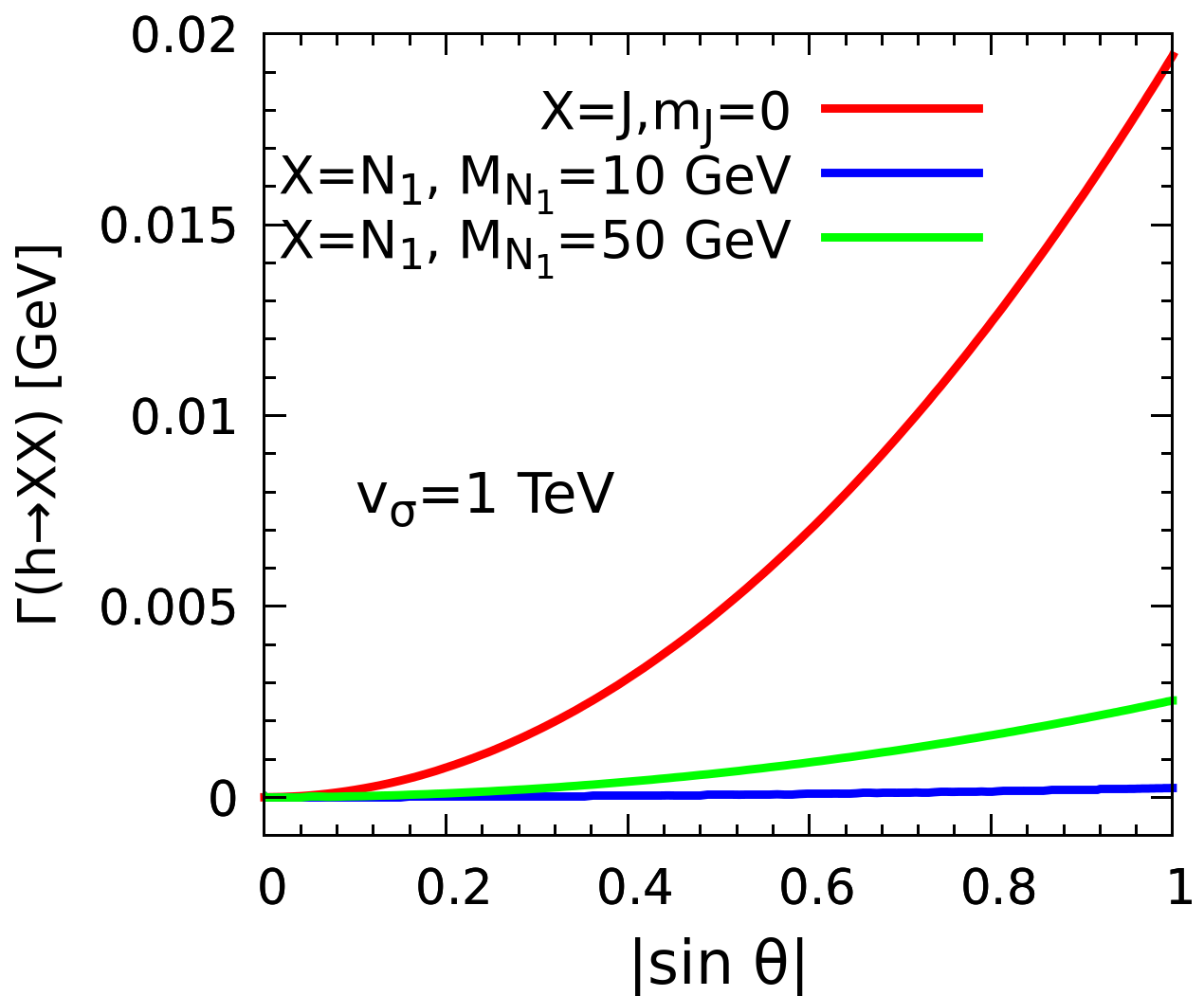}
	\caption{Left panel: The shaded areas on $\sin\theta$ versus $v_\sigma$ are ruled out by the present limit on the invisible Higgs decay $\text{BR}^{\text{inv}}(h) < 0.18$~(magenta) and the signal strength parameter $0.8\leq \mu_f\leq 1$~(gray). See text for details. Right panel: Comparison of decay width $\Gamma(h\to JJ)$ and $\Gamma(h\to N_1 N_1)$ as a function of light-heavy Higgs mixing. Red line stands for $\Gamma(h\to JJ)$, blue and green line stands for $\Gamma(h\to N_1 N_1)$ with $M_{N_1}=10$~GeV and $M_{N_1}=50$~GeV, respectively. We fix the VEV $v_\sigma=1$~TeV.}
	\label{fig:BRINV}
\end{figure}
If the fermionic DM mass $M_{N_1}>m_h/2$, then only the majoron channel $\Gamma(h \to JJ)$ contributes to the invisible Higgs decay. Due to the mixing between light and heavy Higgs, SM like Higgs couplings deviate from the SM values. These will modify the  ``signal strength parameter" $\mu_f$ associated to a given ``visible" final state $f$, which is tested at the LHC~\cite{ATLAS:2020qdt}. Signal strength parameter is defined as follows,
\begin{align}
\mu_f=\frac{\sigma^{\rm NP}(pp\to h) \text{BR}^{\rm NP}(h\to f)}{\sigma^{\rm SM}(pp\to h) \text{BR}^{\rm SM}(h\to f)},
\label{eq:muf}
\end{align}
where $\sigma$ is the cross section for Higgs production and NP stands for the new physics contribution. The available data on $\mu_f$ is separated by production process, see Ref.~\cite{ATLAS:2020qdt}. There are still large uncertainties in these measurements and in view of this here we adopt conservative range $0.8\leq \mu_f\leq 1$.
\par One can translate the invisible Higgs decay and signal strength parameter constraints into the upper bound on the $\sin\theta - v_\sigma$ plane as shown in the left panel of Fig.~\ref{fig:BRINV}. The magenta and gray regions are excluded from the current invisible Higgs decay and $\mu_f$ constraints. As can be seen, from Fig.~\ref{fig:BRINV}, for relatively lower values of $v_\sigma$~(upto 1 TeV), both constraint lead to similar limits on $\sin\theta$. However, for $v_\sigma>1$~TeV, the limit from invisible branching is relatively  relaxed. This can be understood from the fact that larger the $v_\sigma$, the smaller the invisible decay mode $h\to JJ$. Note that these constraints are valid if the fermionic DM mass $M_{N_1}>m_h/2$. However, we find that even in the case of $M_{N_1}<m_h/2$, the exclusion region does not change much. This can be understood from the right panel of Fig.~\ref{fig:BRINV}, where we compare the decay mode $\Gamma(h \to JJ)$ with $\Gamma(h \to N_1 N_1)$ for two values of $N_1$ mass, $M_{N_1}=10$~GeV and $M_{N_1}=50$~GeV. We clearly see that invisible decay width $\Gamma(h \to JJ)$ always dominates over the decay width $\Gamma(h \to N_1 N_1)$.
\par Note also that there will be additional contribution to the diphoton decay channel $h\to\gamma\gamma$ as the SM Higgs boson $h$ also couples to the charged Higgs $\eta^\pm$~\footnote{Note that this charged scalar contributions to $h\to\gamma\gamma$ are generic features of inert doublet schemes~\cite{Cao:2007rm} as well as Scotogenic models~\cite{Ma:2006km}.}. The analytical formula for $h\to\gamma\gamma$ decay width including this new contribution is given in Ref.~\cite{Barbieri:2006dq,Cao:2007rm}. We define the following parameter to measure the deviation from the SM prediction:
\begin{align}
R_{\gamma\gamma}=\frac{\text{BR}(h\to\gamma\gamma)}{\text{BR}(h\to\gamma\gamma)^{\rm SM}}.
\label{eq:Rgammagamma}
\end{align}
The value we use for the SM is $\text{BR}(h\to\gamma\gamma)^{\rm SM}\approx 2.27\times 10^{-3}$. Unlike 8 TeV data, there is currently no combined final data for the 13 TeV Run-2, and the data that is available is separated by the production process~\cite{ATLAS:2019nkf}. In view of this we choose to use 13 TeV ATLAS result which gives the global signal strength measurement of $R_{\gamma\gamma}^{\rm exp}=0.99_{-0.14}^{+0.15}$~\cite{ATLAS:2018hxb}. We find that charged Higgs mass values $m_{\eta^\pm}>100$~GeV can not be ruled out from the experimental limits on $R_{\gamma\gamma}$ from ATLAS~\cite{ATLAS:2018hxb}.
\par Before we conclude this section, we should mention that there are additional limits from the LEP-I and LEP-II experiments. The precise LEP-I measurements rule out SM-gauge bosons decays
to inert scalar particles~\cite{Cao:2007rm,Lundstrom:2008ai}, hence this imposes the following restrictions:
\begin{align}
m_{\eta^R}+m_{\eta^I},2m_{\eta^{\pm}}>m_Z,\,\,\text{and}\,\,m_{\eta^R/\eta^I}+m_{\eta^{\pm}}> m_{W}.
\end{align}
On the other hand, there is no dedicated analysis of LEP-II data in the context of Scotogenic or inert doublet models~(IDM). However, we find that Ref.~\cite{Lundstrom:2008ai}
has discussed LEP-II limits for the case of IDM. Roughly speaking, their LEP II analysis exclude models, leading to the limits satisfying
\begin{align}
 m_{\eta^R} < 80 \text{ GeV},\,\, m_{\eta^I} < 100 \text{ GeV}\,\, \rm{and} \,\, \Delta m \equiv m_{\eta^I}-m_{\eta^R}>8\text{ GeV}.
 \end{align}
Note that LEP also exclude the charged scalar mass  $m_{\eta^\pm} > m_W$~\cite{Pierce:2007ut}.
\begin{figure}[htb!]
	\centering
	\includegraphics[width=0.4\textwidth]{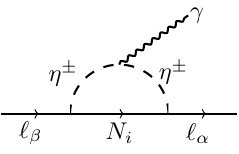}~~~~
	\includegraphics[width=0.4\textwidth]{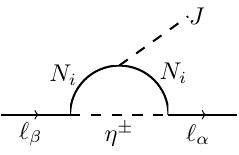}
	\caption{Feynman diagrams for lepton flavor violating processes  $\ell_\beta \to \ell_\alpha\gamma$ and $\ell_\beta \to \ell_\alpha J$ in Dynamic Scotogenic Model.}
	\label{fig:LFV_diag}
\end{figure}

\section{Charged Lepton Flavor Violation and Astrophysical Constraints}
\label{sec:cLFVandStellar}
\subsection{Charged Lepton Flavor Violation}
\label{subsec:cLFV}
The Yukawa interaction $Y^\nu \bar{L}\tilde{\eta} N$ is not only responsible for neutrino mass generation, but they also give rise to cLFV process $\ell_\beta\to\ell_\alpha\gamma$, see the left panel of Fig.~\ref{fig:LFV_diag}. The branching ratio of $\ell_\beta \to \ell_\alpha\gamma$  has the following expression~\cite{Toma:2013zsa,Lindner:2016bgg}:
\begin{align}
\text{Br}(\ell_\beta \to \ell_\alpha\gamma)=\frac{3 \alpha_{\rm em}}{64\pi G_{F}^{2}m_{\eta^\pm}^{4}}\Big|\sum_{i=1}^{3}Y^{\nu}_{\alpha i}Y^{\nu *}_{\beta i}F_{2}(M_{N_{i}}^{2}/m_{\eta^\pm}^{2})\Big|^2 \text{Br}(\ell_\beta \to \ell_\alpha \nu_{\beta}\overline{\nu_{\alpha}})
\label{eq:BRmue2gamma}
\end{align}
where $\alpha_{\rm em}=e^2/4\pi$ and $G_F$ is the Fermi constant. The loop function $F_{2}(x)$ is given by~\cite{Toma:2013zsa},
\begin{align}
F_{2}(x)=\frac{1-6x+3x^{2}+2x^{3}-6x^2\ln x}{6(1-x)^4}
\label{eq:BRmue2gammaLoop}
\end{align}
In this model, one can also have cLFV process such as $\ell_\beta \to \ell_\alpha J$ due to the one-loop coupling of majoron to pair of charged leptons, see right panel of Fig.~\ref{fig:LFV_diag}. The resulting effective vertex $J-\ell_\beta-\ell_\alpha$ can be written as~\cite{Escribano:2020wua},
\begin{align}
\mathcal{L}_{Jl_{\alpha}l_{\beta}}=J\bar{l}_{\beta}(S_{L}^{\beta\alpha}P_{L}+S_{R}^{\beta\alpha}P_{R})l_{\alpha}+h.c.
\label{eq:majoronInteration}
\end{align}
where $P_{L,R}=\frac{1}{2}(1\mp\gamma_{5})$ are the chiral projection operators and $S_{L,R}$ are given by,
\begin{align}
S_{L}=\frac{-i}{16\pi^{2}v_{\sigma}}M_{l}Y^{\nu\dagger}\Gamma Y^{\nu},\,\,\,\,\,
S_{R}=\frac{i}{16\pi^{2}v_{\sigma}}Y^{\nu\dagger}\Gamma Y^{\nu}M_{l}.
\label{eq:mue2J1}
\end{align}
In the above,  $M_{l}$=diag$(m_{e},m_{\mu},m_{\tau})$ and $\Gamma_{mn}$ is defined as,
\begin{align}
\Gamma_{mn}=\frac{M_{N_{n}}^{2}}{(M_{N_{n}}^{2}-m_{\eta}^{2})^{2}}(M_{N_{n}}^{2}-m_{\eta}^{2}+m_{\eta}^{2}\log (m_{\eta}^{2}/M_{N_{n}}^{2}))\delta_{mn}.
\label{eq:mue2J2}
\end{align}
The partial decay width of process $\ell_\beta \to \ell_\alpha J$ is given by,
\begin{align}
\Gamma(\ell_\beta \to \ell_\alpha J)=\frac{m_{\beta}}{32 \pi}(|S^{\beta\alpha}_{L}|^{2}+|S^{\beta\alpha}_{R}|^{2}).
\label{eq:Wmue2J}
\end{align}

\begin{figure}[]
	\centering
	\includegraphics[width=0.7\textwidth]{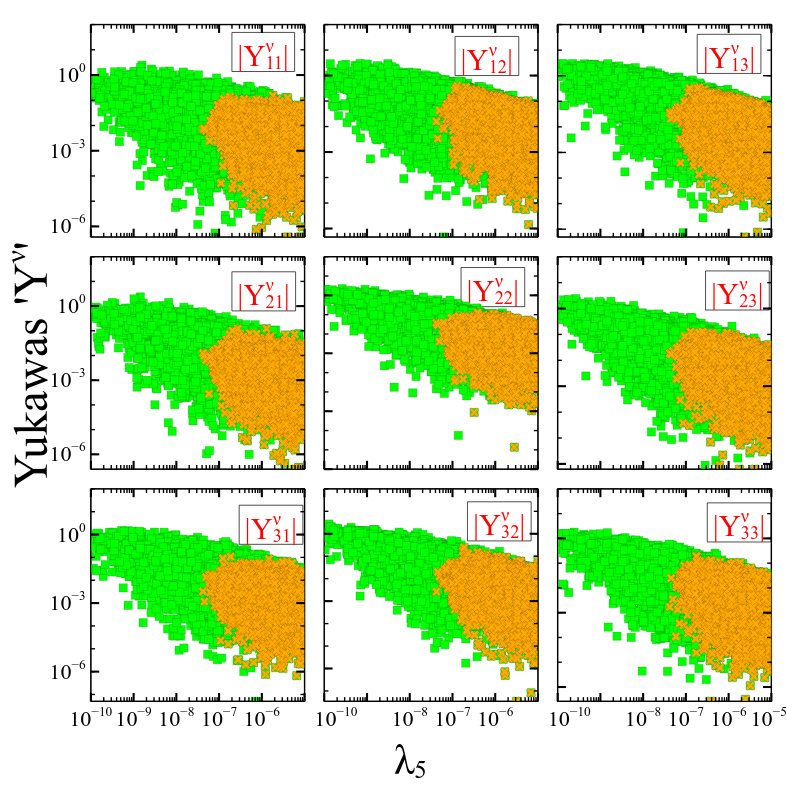}
	\caption{The values of different elements of Yukawa couplings $Y^{\nu}$. The orange points are allowed from $\mu \to e\gamma$, $\mu \to eJ$ and astrophysical constraints corresponding to Dynamic Scotogenic model whereas green points are allowed points from $\mu \to e\gamma$ for Vanilla Scotogenic model. All the allowed parameters also satisfy the best-fit  neutrino oscillation data.}
	\label{fig:yukawa_struct}
\end{figure}

The current experimental upper bound for these processes are \cite{MEG:2016leq,ParticleDataGroup:2016lqr,TWIST:2014ymv},
\begin{align}
\text{Br}(\mu \to e\gamma)\leq 4.3\times10^{-13},\,\,
\text{Br}(\tau \to \mu\  (e)\gamma)\leq 4.4\ (3.3)\times10^{-8},\,\,
\text{Br}(\mu \to e J)\leq 10^{-5}.
\label{eq:lfv_expt_bound}
\end{align}
\begin{table}[]
	\centering
	\begin{tabular}{| c | c  | }
		\hline
		{Parameter}& \multicolumn{0}{ c |   }{\quad Scanned range }   \\
		\hline
		$m_{H}$ [GeV] & \quad [$500$ , $1500$]         \\                                                                            
		$m_{\eta}$ [GeV]  & \quad  [$10^{2}$ , $10^{4}$]                 \\
		$v_{\sigma}$ [GeV]        &  \quad [$500$ , $10^{4}$]                 \\
		$\lambda_{5}$       &   \quad [$10^{-11}$ , $10^{-3}$]          \\
		$Y_{N_{1,2,3}}$       &   \quad [$0.1$ , $2$]                \\
		$\sin \theta$        & \quad   [$10^{-3}$ , $10^{-1}$]  \\
		$\theta_{x,y,z}^{R}$        & \quad  [$0$ , $2\pi$]                   \\
		$\theta_{x,y,z}^{I}$        & \quad  [$0$ , $2\pi$]                    \\
		\hline
	\end{tabular}
	\caption{\centering Input parameters used in our numerical scan to determine allowed parameter space for our model.}
	\label{Table:Scan}
\end{table}
\begin{figure}[htb!]
	\centering
	\includegraphics[width=0.40\textwidth]{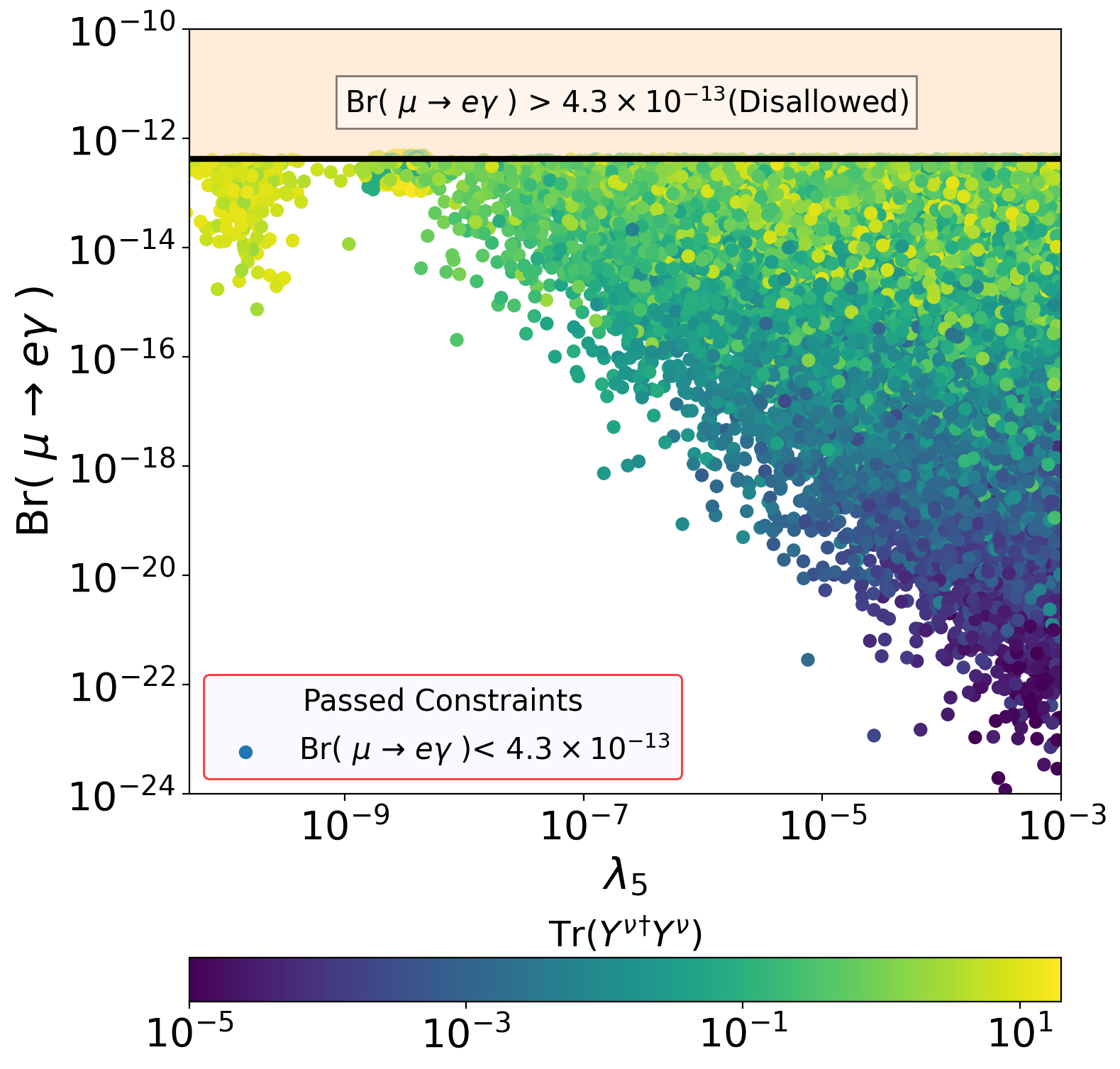}
	\includegraphics[width=0.40\textwidth]{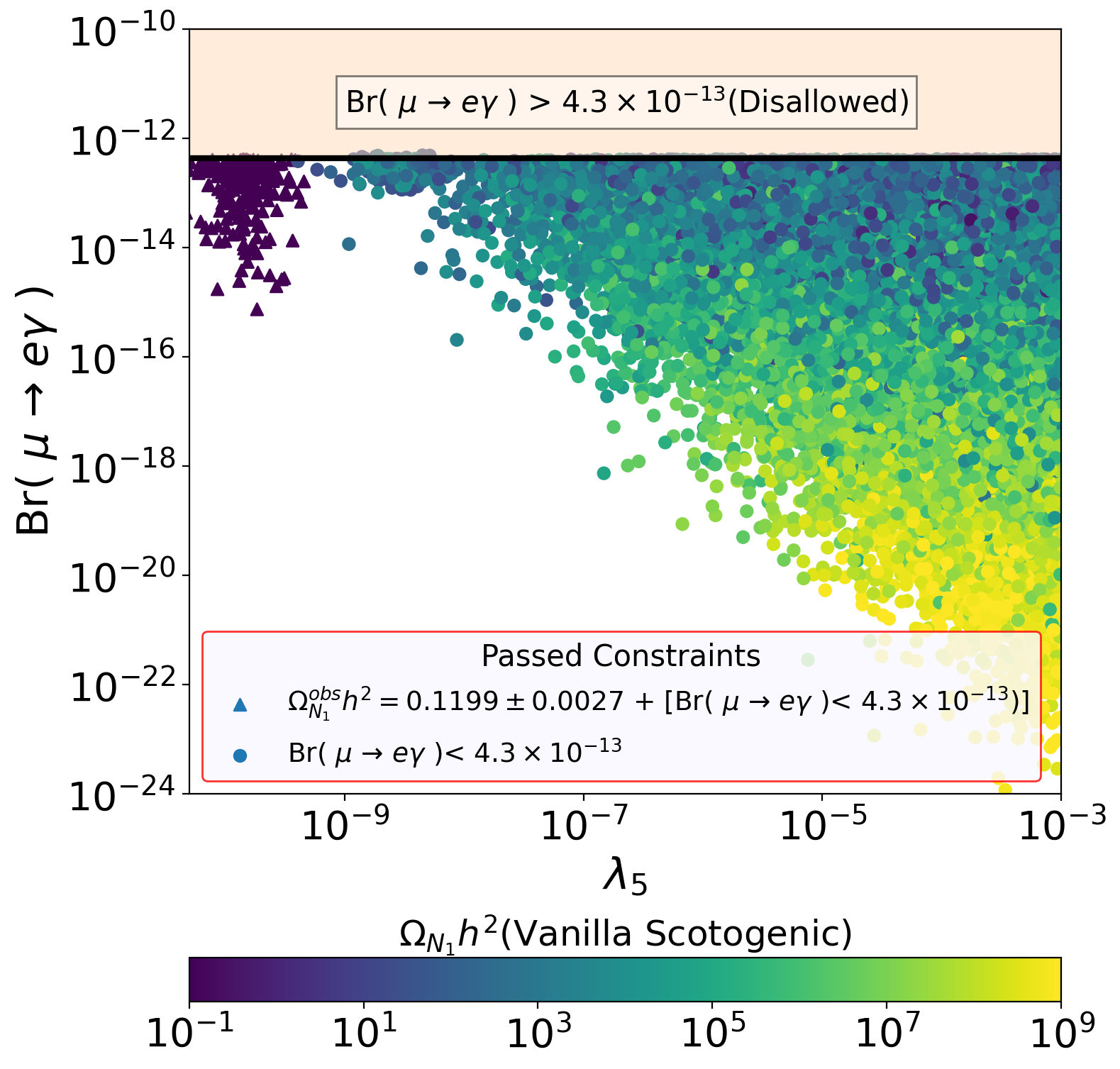}
	\includegraphics[width=0.40\textwidth]{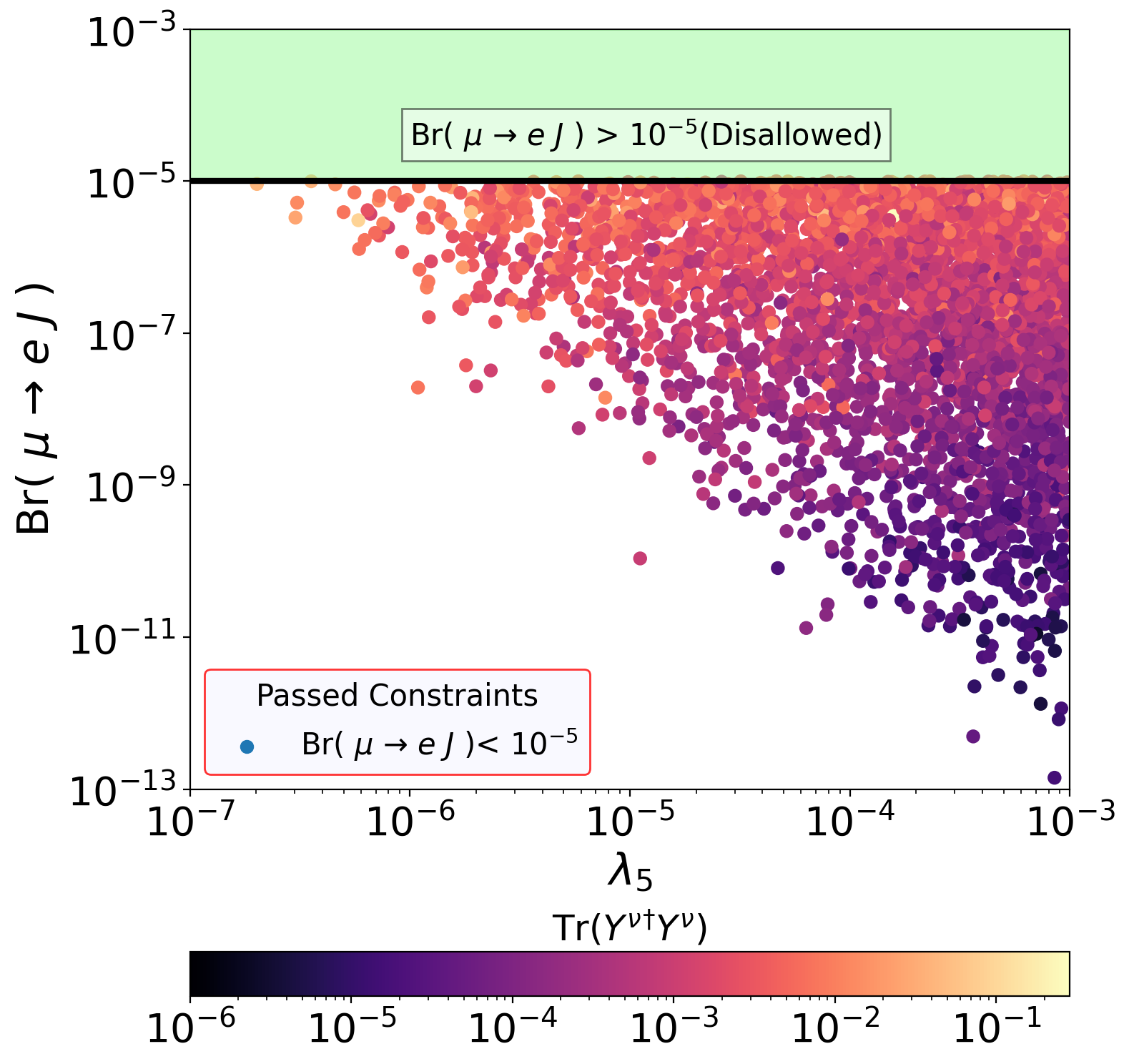}
	\includegraphics[width=0.40\textwidth]{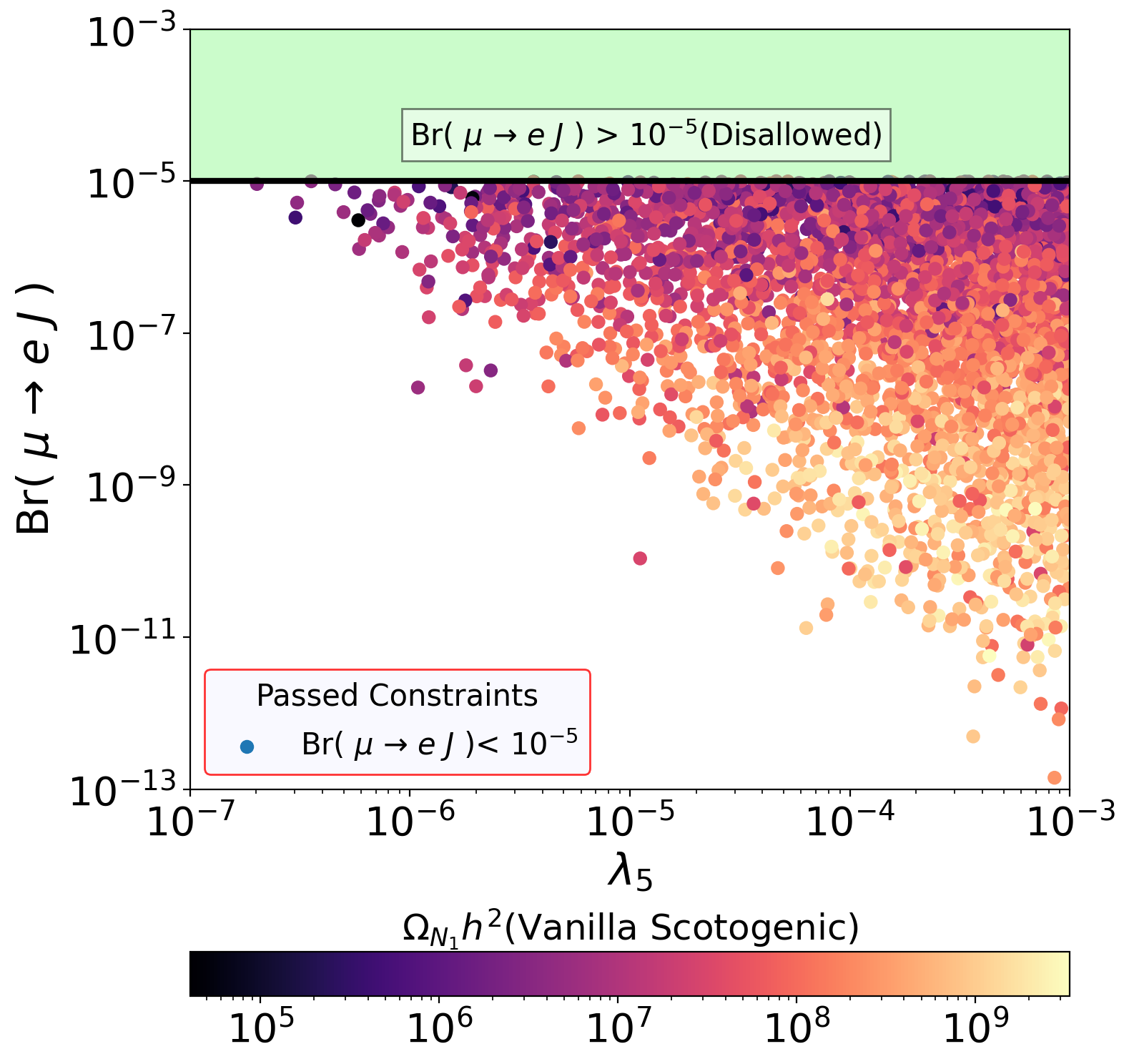}
	\caption{Upper and lower left panels: variation of the branching ratios of $\mu \to e\gamma$ and $\mu \to e J$ processes w.r.t the coupling $\lambda_5$. The color panel represents {Tr}$({Y^{\nu}}^{\dagger}Y^{\nu})$. Upper and lower right panel: the same, as the left panel, however the color panel represents the contribution $\Omega_{N_1}h^2$ in the DM relic density, which appears from the vanilla Scotogenic diagram $N_{1} N_{1} \to \ell^{+} \ell^{-}/ \nu_\ell \nu_\ell$.  }
	\label{fig:Br-LFV1}
\end{figure}
Since these diagrams in Fig.~\ref{fig:LFV_diag} are proportional to the Yukawa coupling $Y^{\nu}$, hence a large $Y^{\nu}$ is severely constrained from these above cLFV constraints. We find that the stringent bound for our model arises from $\mu \to e\gamma$ and $\mu \to e J$ processes respectively. To do this, we varied the model parameters in the range shown Table.~\ref{Table:Scan} while keeping $\lambda_3 =\lambda_{4} =0.01$, $\lambda_{\eta \sigma}=0.001$ fixed. The Yukawa coupling $Y^\nu$ is determined by Eq.~\ref{eq:Ynu} where we fix the oscillations parameters to their best fit values~\cite{deSalas:2020pgw}. Also in our analysis, $M_{N_1}< m_{\eta^{R,I}}, m_{\eta^\pm}, M_{N_{2,3}}$ has been taken to make sure that $N_1$ is indeed the lightest dark sector particle. In the upper left panel of Fig.~\ref{fig:Br-LFV1}, we show the variation of the branching ratio of $\mu \to e \gamma$ with respect to the coupling $\lambda_5$. The colour pallet represents $\rm{Tr}({Y^{\nu}}^{\dagger}Y^{\nu})$. From Eq.~\ref{eq:numass2}, we see that Yukawa coupling $Y^\nu$ can be large for small $\lambda_5$. This can also be seen from Fig.~\ref{fig:yukawa_struct}, where we have shown the variation of different matrix elements of $Y^{\nu}$ with respect to $\lambda_{5}$ satisfying $\text{BR}(\mu\to e\gamma)$ constraints. The green points clearly shows that large Yukawa couplings are allowed but only for small $\lambda_5$. If the mass of DM $N_1$ is far away from $\eta$ and $N_{2,3}$, the only annihilation channels in the pure Scotogenic model which determines the DM relic abundance are $N_1 N_1\to \ell^+\ell^-/\nu_\ell\nu_\ell$ via the Yukawa coupling $Y^\nu$. This diagram is mediated via $t$-channel propagation of $\eta^{\pm}, \eta^{R/I}$, and referred as ``vanilla Scotogenic" diagram. The analytical expression for thermal average annihilation cross section is given in Ref.~\cite{Kubo:2006yx} and can be written as
\begin{align}
\langle \sigma v \rangle_{N_{1}N_{1}\to \ell^{+}\ell^{-},\nu\nu}=\frac{6r_{1}^{2}(1-2r_{1}+2r_{1}^{2})\sum_{i,j}|Y^{\nu}_{i1} Y^{\nu *}_{j1}|^{2} }{24\pi M_{N_{1}}^{2}x_{f}},\hspace{0.51cm} r_1= M_{N_{1}}^{2}/(M_{N_{1}}^{2}+m_{\eta}^{2})
\label{eq:PAn1s}
\end{align}
The relic density is then given as
\begin{equation}
\begin{aligned}
\Omega_{N_{1}}h^{2}=1.756\times 10^{-11}\left(\frac{\langle \sigma v \rangle_{N_{1}N_{1}\to \ell^{+}\ell^{-},\nu\nu}}{x_{f}} \right)^{-1},
\label{eq:vanillascot}
\end{aligned}
\end{equation}
where $x_f$ is the ratio $M_{N_1}/T_{f}$ at the freeze-out temperature and is given by
\begin{align}
x_{f}=\ln \frac{0.0302 M_{\rm pl} M_{N_{1}}\langle \sigma v \rangle_{N_{1}N_{1}\to \ell^{+}\ell^{-},\nu\nu}}{x_{f}^{1/2}}.
\label{eq:Ftemp}
\end{align}
In the upper right panel of Fig.~\ref{fig:Br-LFV1}, we show the variation of the branching ratio of $\mu \to e \gamma$ vs the coupling $\lambda_5$, where the color panel represents the contribution in the DM relic density, which appears from $N_{1} N_{1} \to \ell^{+} \ell^{-}/ \nu_\ell \nu_\ell$ annihilation. In order to obtain the correct relic abundance, the magnitude of $\lambda_{5}$ is approximately around $10^{-10}$\cite{Vicente:2014wga,Liu:2022byu}. Further decrease in $\lambda_{5}$ is in direct conflict with cLFV constraint\footnote{However, these cLFV constraints can be evaded when co-annihilation effects for DM relic abundance become important.}. From this figure, it is evident, that this diagram alone can satisfy the observed DM relic density only in a restricted parameter space of $\lambda_{5}$ . It is also important to highlight that the points satisfying the observed relic density satisfy the hierarchical Yukawa structure $|Y^{\nu}_{11}|\lsim|Y^{\nu}_{21}|\lsim|Y^{\nu}_{31}|$ as result of the $\mu\to e\gamma$ bound which has been shown in Ref.~\cite{Vicente:2014wga}. Thus, dark matter $N_{1}$ behaves as $\mu\tau$-philic particles and annihilates dominantly into second and third family leptons. Note that the analysis of fermionic dark matter in the scotogenic model without the effect of co-annihilation was done in \cite{Vicente:2014wga} and we find that our results are in complete agreement with their results.

In the lower left panel of Fig.~\ref{fig:Br-LFV1}, we show the variation of  $\mu \to e J$ with respect to $\lambda_5$.  Comparing the upper and lower left panel, it is evident that $\mu \to e J$ imposes much more severe constraint as compared to  $\mu \to e\gamma$ due to their different loop structures. From the figure, it is evident that  the Yukawa coupling $\rm{Tr}({Y^{\nu}}^{\dagger} Y^{\nu}) > 10^{-1}$ is mostly disallowed\footnote{The expected future sensitivity of $\mu \to e J$ process from MEG-II\cite{Jho:2022snj,Calibbi:2020jvd} and Mu3e experiment~\cite{Perrevoort:2018okj} will futher decrease the upper bound on $\rm{Tr}({Y^{\nu}}^{\dagger} Y^{\nu})$ by approximately an order of magnitude, leading to more feeble $Y^{\nu}$ but this will not change the conclusion of the paper as we will show that the correct relic abundance obtained through dark matter annihilation to majoron will be unaffected.}. For this limiting Yukawa value, the relic density from vanilla Scotogenic diagram is even larger $\Omega_{N_1} h^2 > 10^{4}$, which is completely disallowed. Therefore, pure vanilla Scotogenic diagram in this model alone is not adequate to satisfy the experimentally observed DM relic density and additional contributions must be taken into account. In Fig.~\ref{fig:yukawa_struct}, we have illustrated with orange points for our model that the constraints from $\mu \to e\gamma$, $\mu \to e J$ and astrophysical constraints\footnote{We discuss astrophysical constraints from stellar cooling and $\Delta N_{eff}$ in detail for our model in next Sec.~\ref{sec:Astrophysical_Constraints}.} gets severely stringent for $\lambda_{5}<10^{-7}$. Indeed as we will discuss in Sec.~\ref{sec:DManalysis}, one can easily obtain correct relic density with the help of additional annihilation channels involving majoron~($J$) and second Higgs~($H$) even in absence of co-annihilation effects.
\subsection{Astrophysical Constraints}
\label{sec:Astrophysical_Constraints}
Due to the presence of a massless majoron state, the model is subject to different  astrophysical constraint, such as, stellar cooling of compact stars. Additionally, massless majoron being relativistic species can alters $N_{\rm eff}$~\cite{Weinberg:2013kea}. Below, we discuss them in detail. 
\begin{figure}[]
	\centering
	\includegraphics[width=0.49\textwidth]{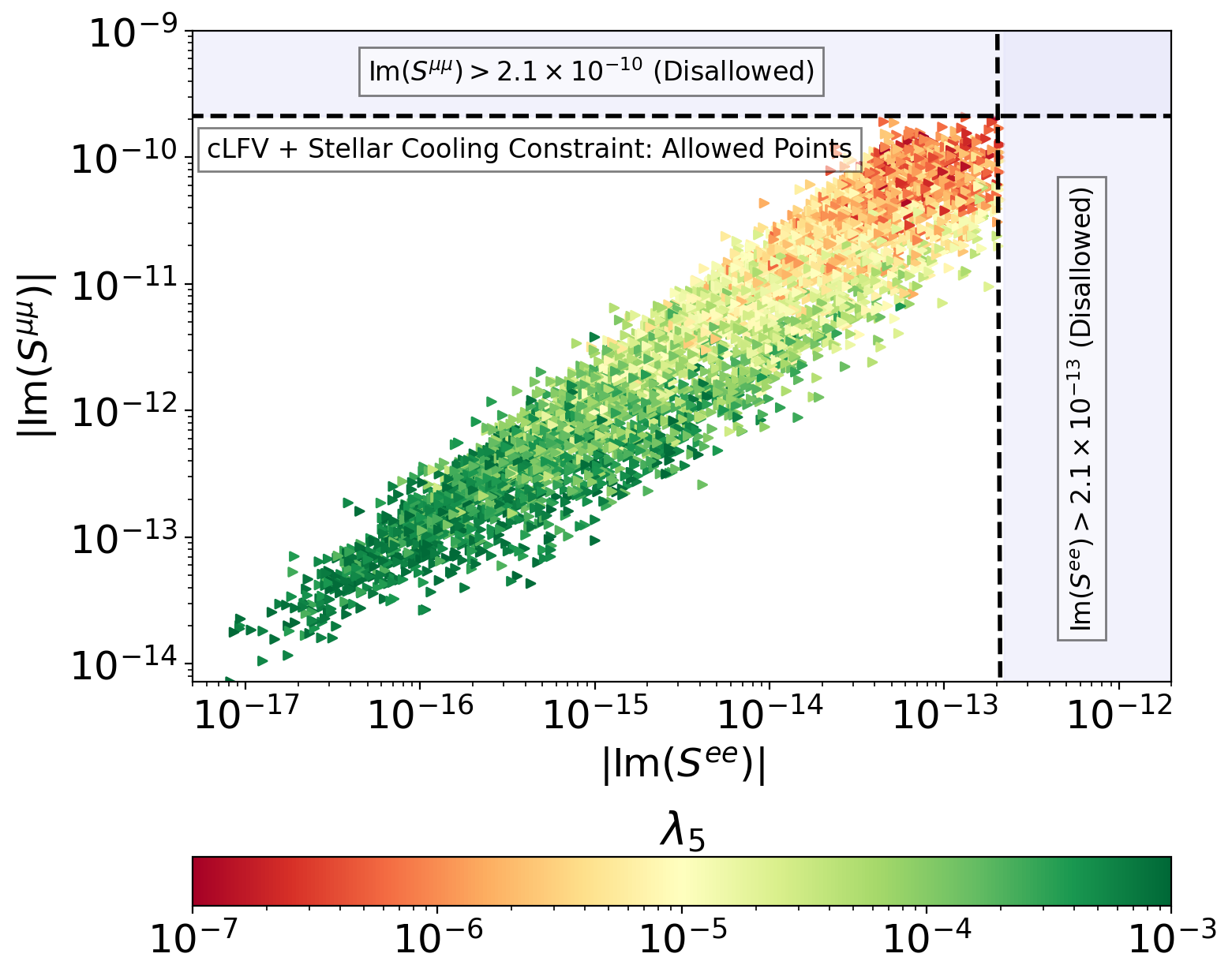}
	\caption{We show allowed parameter space from stellar cooling+$\text{BR}(\mu \to e\gamma)+\text{BR}(\mu \to e J)$ with colored pallet corresponding to allowed value of $\lambda_{5}$. All the allowed points also satify the best fit neutrino oscillation data.}
	\label{fig:astro-diagrams1}
\end{figure}
\subsubsection{Stellar Cooling}
The massless relativistic  particle  majoron gets stringent constraints from the stellar objects such as  Red Giants (RD), White Dwarafs (WD) and Horizontal Stars (HS). If the majoron is  massless or is in the KeV range, then it can be produced copiously inside the core of the stellar objects~\cite{Dev:2020jkh,Escribano:2020wua}, which in turn may lead to efficient stellar cooling by transporting energy from the stellar core of the compact star provided that majoron interacts weakly with stellar medium. If the stellar cooling takes places at high rate, it can alter the observed luminosity of stellar objects leading to conflict with astrophysical observations. Thus, the stellar cooling puts bound on the couplings between majoron and SM particles. For our model, the majoron $J$ can be efficiently produced in the core of stellar objects through process such as, $\ell+\gamma \to \ell + J$ (Compton like), $\ell+ N \to \ell + N + J$ ($\ell$ - N bremsstralung) and $\ell+ \ell \to \ell + \ell + J$ ($\ell$ - $\ell$ bremsstrahlung). The relative importance of each of these processes depends on the density and temperature of the medium, and therefore on the astrophysical scenario. Many studies have recently examined the issue of cooling in astrophysical objects caused by the emission of ultralight pseudoscalars~\cite{DiLuzio:2020wdo,Calibbi:2020jvd,Raffelt:1994ry,Bollig:2020xdr,Croon:2020lrf,DiLuzio:2020jjp}, which are also applicable for majoron. To derive the bounds on couplings for our model, we can rewrite Eq.~\ref{eq:majoronInteration} for $J-\ell_\alpha-\ell_\alpha$ as below,
\begin{align}
\mathcal{L}^{\rm diag}_{J\ell_{\alpha}\ell_{\alpha}}=-i J\bar{\ell}_{\alpha}\text{Im}(S^{\alpha\alpha})\gamma_{5}\ell_{\alpha},
\label{eq:majoronInteration2}
\end{align}
where $S^{\alpha\alpha}$($=S_{L}^{\alpha\alpha}+S_{R}^{\alpha\alpha *}$) couplings are purely imaginary due to the fact that majoron $J$ being pseudo scalar state. The $S^{\alpha\alpha}$ couplings are related to the model parameters by Eq.~\ref{eq:mue2J1}. The stringent constraint on the couplings of majoron with electrons comes from white dwarfs. The bremsstrahlung process is very efficient in the dense core of a white dwarf. The authors in Ref.~\cite{Calibbi:2020jvd} sets the bound on $\text{Im}\  S^{e e}$ using Sloan Digital Sky Survey and the SuperCOSMOS Sky Survey data,
\begin{align}
\text{Im}\  S^{e e}<2.1\times10^{-13}.
\label{eq:majoronbound1}
\end{align}
Recently the Refs.~\cite{Croon:2020lrf,Calibbi:2020jvd,Bollig:2020xdr,Caputo:2021rux} also studied the supernova SN1987 to put bound on muon coupling to majoron.  In this case the process that has been ultimately used to set the constraint is $\mu+\gamma\to\mu +J$~\cite{Croon:2020lrf},
\begin{align}
\text{Im}\  S^{\mu \mu}<2.1\times10^{-10}.
\label{eq:majoronbound2}
\end{align}
It is clear from Eq.~\ref{eq:majoronbound1} and Eq.~\ref{eq:majoronbound2} that the large couplings of majoron with electron and muon is excluded from stellar cooling and supernova. In  Fig.~\ref{fig:astro-diagrams1}, we show the constraint on $\text{Im}\  S^{\mu \mu}$ and $\text{Im}\  S^{e e}$ from  $\mu \to e \gamma$, $\mu \to eJ$ and from stellar cooling. It can be seen the allowed parameter space from stellar cooling, $\text{BR}(\mu \to e\gamma)$ and $\text{BR}(\mu \to e J)$ broaden with increase in $\lambda_{5}$ coupling strength. The allowed parameters  also satisfy the best-fit  neutrino oscillation data. 
\begin{figure}[]
	\centering
	\includegraphics[width=0.3\textwidth]{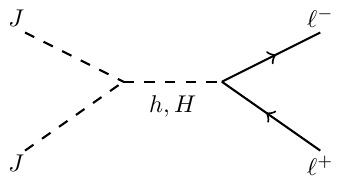}
	\includegraphics[width=0.25\textwidth]{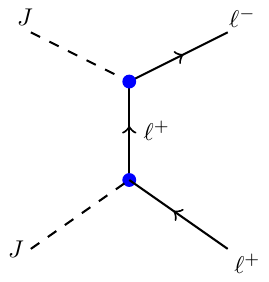}
	\includegraphics[width=0.25\textwidth]{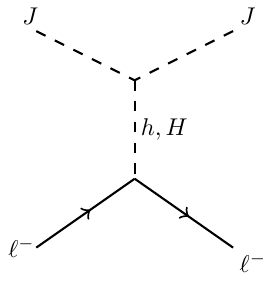}
	\caption{Diagrams for the majoron annihilation and scattering relevant for the chemical and kinetic equillibrium with thermal bath. The blue dots represent the effective $J-\ell-\ell$ vertex. }
	\label{fig:neff}
\end{figure}
\subsubsection{$\Delta N_{\rm eff}$}
The majoron $J$ is associated with breaking of spontaneously global continuous symmetry $U(1)_{B-L}$. It was pointed in \cite{Weinberg:2013kea} that majoron $J$ may mimic as an  additional neutrino species in measurements of the anisotropies in the cosmic microwave background (CMB) if majoron remain in thermal equilibrium with ordinary particles until after the era of muon annihilation. The effective number of neutrino species that existed prior to recombination era, $N_{\rm eff}$, can be used to measure this effect.  The contribution of majoron to $N_{\rm eff}$ can be determined by evaluating the decoupling temperature of majoron $T_{d}^{J}$. $\Delta N_{\rm eff}$ can be written as \cite{Garcia-Cely:2013nin,Bernal:2015bla},
\begin{align}
\Delta N_{\rm eff}=N_{\rm eff}-3=\frac{4}{7}\left[\frac{g(T_{d}^{\nu})}{g(T_{d}^{J})}\right]^{4/3}
\label{eq:Neff},
\end{align}
where $g(T)$ is the relativistic degree of freedom at given temperature $T$ and $T_{d}^{\nu}$ corresponds to the decoupling temperature of neutrinos from the thermal bath. The majoron decoupling temperature $T_{d}^{J}$ is determined through this relation,
\begin{align}
\left(\frac{n_{J}^{eq}\sum_{f}\langle \sigma v \rangle_{J J \to \bar{f}f}}{H(T)}\right)_{T=T_{d}^{J}}=1
\label{eq:TdJ},
\end{align}
where $H$ is the Hubble expansion rate, $n_{J}^{eq}$ is equilibrium number density and $\langle \sigma v \rangle_{J J \to \bar{f}f}$ is thermal average cross section which is given by,
\begin{align}
\langle \sigma v \rangle_{J J \to \bar{f}f}=\frac{1}{32 T^{5}}\int_{4 m_{f}^{2}}^{\infty}\sigma(JJ\to\bar{f}f)s\sqrt{s}K_{1}(\sqrt{s}/T)ds
\label{eq:sigmavJ1},
\end{align}
with,
\begin{align}
\sigma(JJ\to\bar{f}f)=\frac{m_{f}^{2}\lambda_{\Phi\sigma}^{2}}{8\pi}\frac{(1-4 m_{f}^{2}/s)^{3/2}(s^{2}(m_{h}-m_{H})^2+m_{H}^2 m_{h}^{2}(m_{H}\Gamma_{h}-m_{h}\Gamma_{H})^{2})}{(m_{h}-m_{H})^2((s-m_{h}^{2})^{2}+\Gamma_{h}^{2}m_{h}^{2})((s-m_{H}^{2})^{2}+\Gamma_{H}^{2}m_{H}^{2})}
\label{eq:sigmavJ2},
\end{align}
where $\Gamma_{H,h}$ is total decay width of BSM and SM Higgs. In Fig.~\ref{fig:neff}, we show the relevant processes that keeps majoron $J$ in chemical and kinetic equillibrium with thermal bath in early epoch of the universe. We have assumed during freeze-out of majoron $J$, the majoron is kept in kinetic equillibrium by frequent elastic scattering with relativistic SM particles. The first two diagrams contributes to number changing process of majoron $J$. On the other hand, the last diagram keeps the majoron in kinetic eqillibrium with thermal bath. In the second Feynman diagram, $ J J \to l^{+}l^{-}$, the vertices are loop-suppressed. Further, as we are considering relatively large light-heavy higgs mixing for our dark matter analysis, we find that the contribution from the first diagram are the dominant one in determination  $ \langle \sigma v \rangle_{J J \to \bar{f}f}$.
\begin{figure}[]
	\centering
	\includegraphics[width=0.49\textwidth]{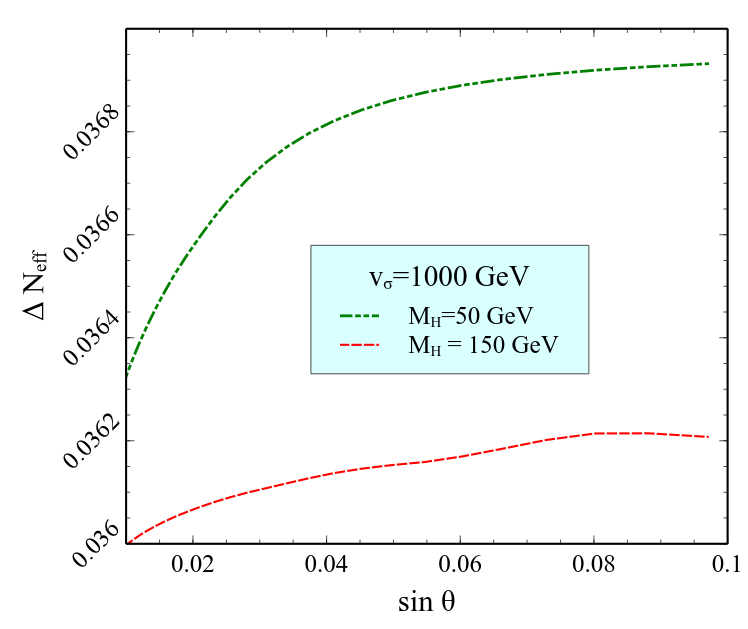}
	\caption{Variation of $\Delta N_{\rm eff}$ vs the Higgs mixing angle $\sin \theta$ for different values of second Higgs mass $m_{H}=50$~GeV and 150 GeV.  It is evident  that $\Delta N_{eff}<0.04$ for our choosen benchmark point. Thus, the majoron decouples from thermal bath early enough to get diluted.}
	\label{fig:astro-diagrams2}
\end{figure}

In Fig.~\ref{fig:astro-diagrams2}, we show how $\Delta N_{\rm eff}$ varies with mixing $\sin\theta$ for two different Higgs mass $m_{H}=50$~GeV and 150 GeV, respectively. We have found that the contribution to $N_{\rm eff}$ is quite negligible as long as $m_H$ is relatively high.
\begin{figure}[]
	\centering
	\includegraphics[width=0.35\textwidth]{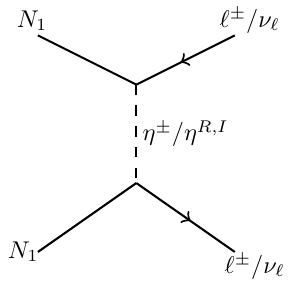}
	\includegraphics[width=0.35\textwidth]{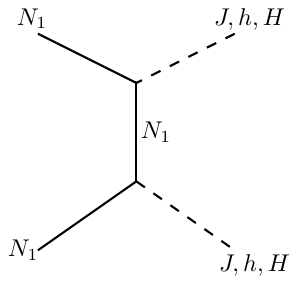}
	\includegraphics[width=0.35\textwidth]{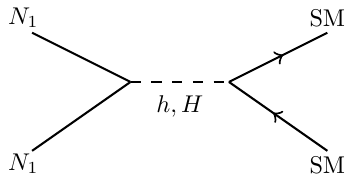}
	\includegraphics[width=0.35\textwidth]{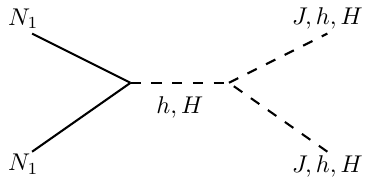}
	\caption{\centering Feynmann Diagram for the dark matter annihilation channels in dynamical Scotogenic model.}
	\label{fig:annihilation-diagrams}
\end{figure}
\begin{table}
	\centering
	\renewcommand\arraystretch{1.5}
	\begin{tabular}{| c | c | c | c | c | }
		\hline 
		\multicolumn{2}{|c|}{Initial state} & \multicolumn{2}{|c|}{Final state} & Scaling with couplings \\
		\hline
		\multirow{2}{*}{$N_{1}$} & \multirow{2}{*}{$N_{1}$} & $l^-$ & $l^+$ & \multirow{2}{*}{$Y^{\nu}{}^4$} \\
		\cline{3-3} \cline{4-4} 
		& & $\nu$& $\bar{\nu}$ &{\bf 't' channel processes} \\
		\hline
		\multirow{3}{*}{$N_{1}$} & \multirow{3}{*}{$N_{1}$} & $J$ & $J$ & \multirow{3}{*}{$Y^{N}{}^4$} \\
		\cline{3-3}\cline{4-4}
		& & $h, H$& $h, H$ & \\
		\cline{3-3}\cline{4-4}
		& & $J$& $h, H$ &{\bf 't' channel processes} \\
		\hline
		\multirow{4}{*}{$N_{1}$} & \multirow{4}{*}{$N_{1}$} & $Z,W^{+}$ & $Z,W^{-}$ & \multirow{4}{*}{$Y^{N}{}^2$} \\
		\cline{3-3}\cline{4-4}
		& & $q$& $q$ & \\
		\cline{3-3}\cline{4-4}
		& & $h,H$& $h, H$ & \\
		\cline{3-3}\cline{4-4}
		& & $J$& $J$ &{\bf 's' channel processes} \\
		\hline
	\end{tabular}
	\caption{List of all relevant annihilation processes with their dependency of the cross sections on the Yukawa coupling strength $Y^{\nu}$ and $Y^{N}$.  }
	\label{tab:ann}
\end{table}
\section{Dark Matter analysis in the dynamical Scotogenic model}
\label{sec:DManalysis}
In Sec.~\ref{sec:cLFVandStellar} we discussed that due to the constraints coming from cLFV, pure vanilla Scotogenic diagram is not adequate to satisfy the observed DM relic density. In this section, we discuss how additional DM annihilation channels involving majoron and second Higgs helps to obtain correct relic density even after taking into account all the relevant constraints mentioned before.
\begin{figure}[]
	\centering
	\includegraphics[width=0.37\textwidth]{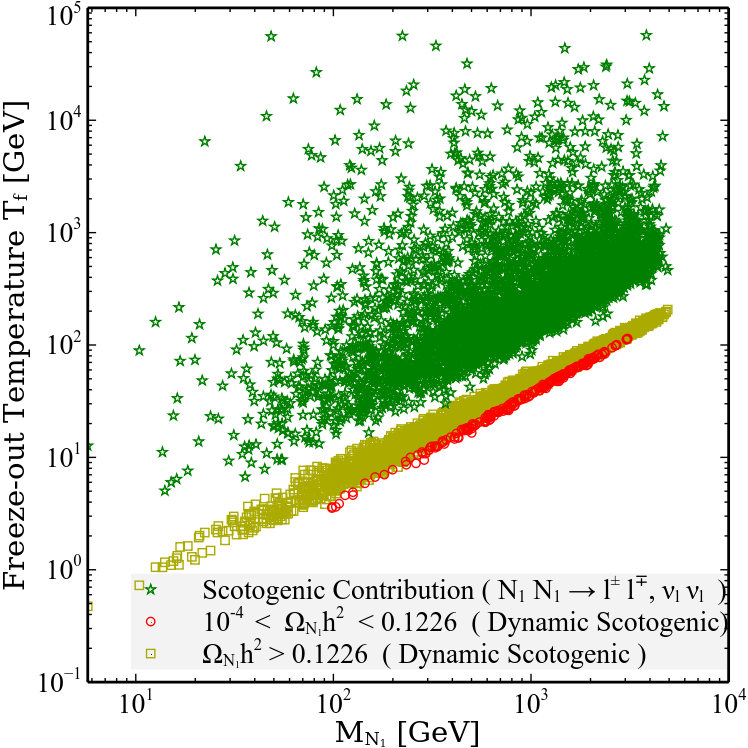}
	\includegraphics[width=0.37\textwidth]{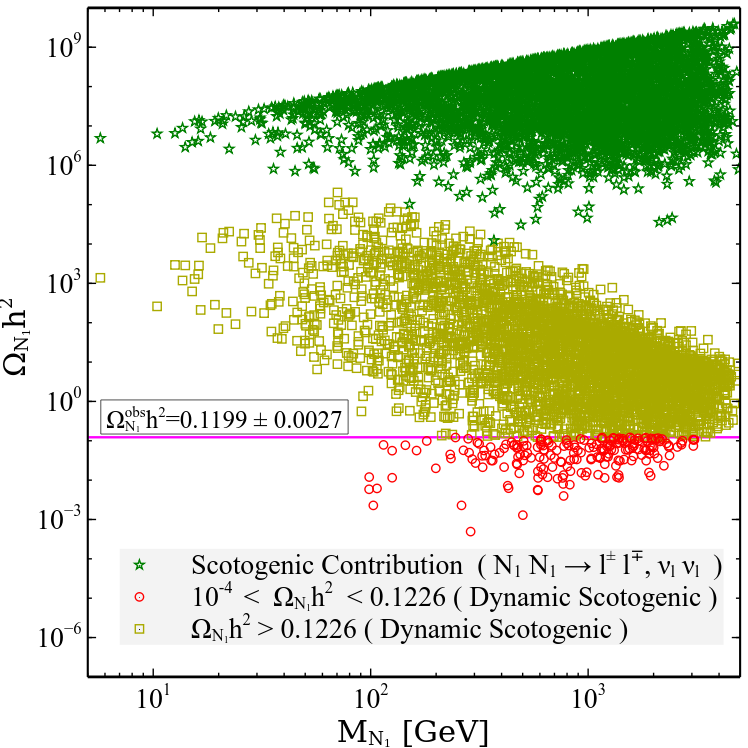}
	\includegraphics[width=0.37\textwidth]{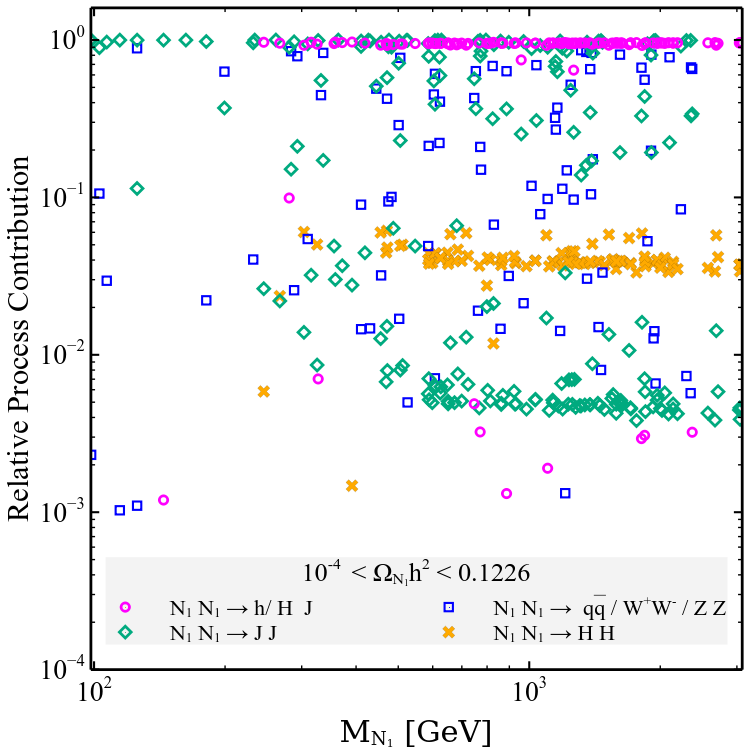}
	\caption{Upper left panel: variation of freeze-out temperature $T_{f}$ with DM mass $M_{N_{1}}$. Upper right panel: the DM relic density $\Omega_{N_{1}}h^{2}$ vs mass of DM $M_{N_{1}}$. Lower panel: variation of relative process contribution with DM mass $M_{N_{1}}$.  }
	\label{fig:Scoto-Annhilation}
\end{figure}

\subsection{Dark matter abundance}
As we have specified before, we consider $N_1$ state as a WIMP DM. The $N_2$, $N_3$ and $\eta^{0}$ states have masses $M_{N_i}(i=2,3),m_{\eta^0}> 1.5M_{N_1}$ which ensures no co-annihilation process with DM \cite{Molinaro:2014lfa}. Hence, the observed relic density of $N_{1}$ is solely attained through DM annihilation. We show  different annihilation diagrams in Fig.~\ref{fig:annihilation-diagrams} and the dependence of 
relevant interaction rates  on Yukawa couplings  in Table.~\ref{tab:ann}. In addition to the $N_1 N_1 \to \ell^{\pm}\ell^{\mp}/\nu_\ell\nu_\ell$ channels, additional channels such as $N_1 N_1 \to J J, h h, H H, J h/H $ and $N_1 N_1 \to \text{SM}\ \text{SM}$ channels are also open. To evaluate the DM  relic abundance, we use MicroOmegas \cite{Belanger:2014vza}. In Fig.~\ref{fig:Scoto-Annhilation}, we 
show  the variation of the relic abundance, freeze-out temperature and relative process contribution vs the mass of the DM $M_{N_1}$.  All the points are in agreement with the cLFV constraint and neutrino oscillation data. In addition, they are also in agreement with the astrophysical constraint discussed in the previous section. It is important to re-emphasise  that the constraint arising from cLFV, particularly $\mu \to e J$ favours somewhat smaller values of $Y^{\nu}$ and large value of $\lambda_{5}$, as is evident from Fig.~\ref{fig:Br-LFV1}. This in turn suppress the cross-section of $N_1 N_1 \to l^{\pm}l^{\mp}/\nu_{l}\nu_{l}$,  as it depends on the fourth power of $Y^{\nu}$.  Because of the suppressed interaction, 
the $N_1 N_1 \to \ell^{\pm}\ell^{\mp}/\nu_\ell\nu_\ell$ process decouples much earlier from the thermal bath compared to other annihilation processes. As can be seen from the upper left panel of Fig.~\ref{fig:Scoto-Annhilation}, the green points corresponding to $N_1 N_1 \to \ell^{\pm}\ell^{\mp}/\nu_\ell\nu_\ell$ process (vanilla Scotogenic diagram) decouples much earlier, thereby leading to a higher  freeze-out temperature $T_f$, that  can be as  large  as $10^4$ GeV for 
the DM mass $M_{N_1} \sim $ TeV.  The DM relic density for these green points are overabundant which is evident from  upper right panel of Fig.~\ref{fig:Scoto-Annhilation}. 

The presence of additional processes such as $N_1 N_1 \to J J, J h/H, h h, H H $ and $N_1 N_1 \to \text{SM}\  \text{SM}$ which are independent of $Y^{\nu}$ but dependent on $Y^{N}$ are 
important to acquire the correct relic density.  This can be realised from the upper right panel of Fig.~\ref{fig:Scoto-Annhilation} when we compare the relic density of $N_{1}$ corresponding to vanilla Scotogenic processes~\footnote{Only after taking into account additional constraints,i.e, $\mu \to e J$ and SCC's , the vanilla scotogenic contribution is overabundant. For a vanilla Scotogenic model, $ \mu \to e J$ constraint is not present. The both observed dark matter relic density and $\mu \to e \gamma$ constraints in Vanilla Scotogenic model get satisfied for small range of $\lambda_{5}$ Ref.~\cite{Vicente:2014wga,Liu:2022byu}.} shown in green points with red and yellow points which represent relic density obtained in the dynamical Scotogenic model after taking into account all the $s$ and $t$ channel processes.  For  the points marked in red, relic 
density varies   in the range between $10^{-4}$ to $0.1226$. In the lower panel of Fig.~\ref{fig:Scoto-Annhilation}, we show the relative contribution of  different channels in the relic abundance. It can be seen, $N_1 N_1 \to J J$ and $N_1 N_1 \to h/H J$ can give large contribution. Other processes, such as  $N_1 N_1 \to\text{SM}\ \text{SM}$ via $s$ channel  mediation is subdominant due to the propagator suppression over majority of parameter space except near $s$ channel resonance region. Taking into account  diagrams relevant for dynamical Scotogenic model, such as $N_1 N_1 \to J J, J h/H$, the DM remains in the thermal equilibrium for a long time, thereby reducing the freeze-out temperature $T_f \sim  1-100$ GeV.  Hence, the DM relic abundance can satisfy the experimental observation.

\subsection{DM Direct detection bound}
\begin{figure}[]
	\centering
	\includegraphics[width=0.35\textwidth]{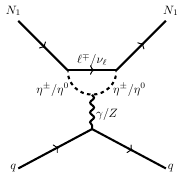}~~~~~~
	\includegraphics[width=0.32\textwidth]{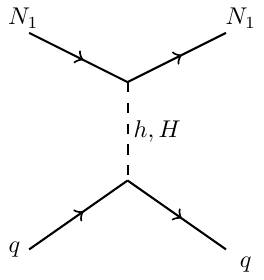}
	\caption{Feynman diagrams for the processes contributing to direct detection in dynamical Scotogenic model. Left panel diagram only contributes in vanilla Scotogenic model.}
	\label{fig:DD-diagram}
\end{figure}
\begin{figure}[t]
	\centering
	\includegraphics[width=0.5\textwidth]{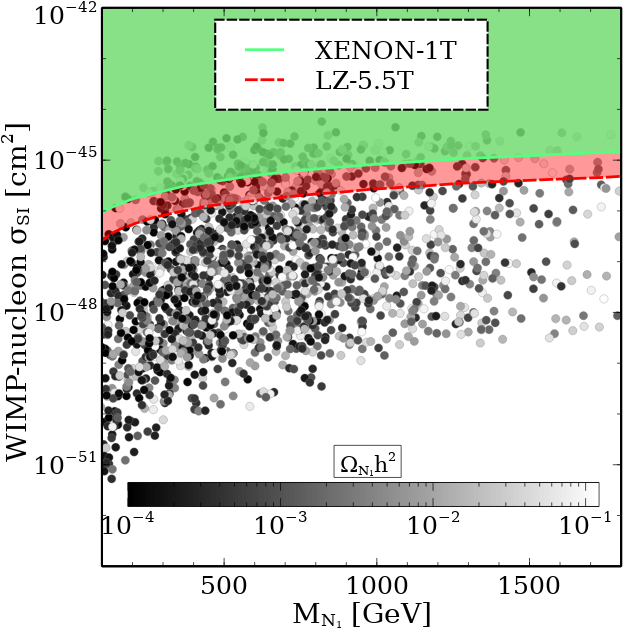}
	\caption{Bounds on the DM mass from direct detection search at the Xenon-1T and LUX-ZEPLIN (LZ) detectors. The colorbar corresponds to the variation of relic density in the range $10^{-4}$ to $0.1226$.}
	\label{fig:DD}
\end{figure}
Let us now study the direct detection prospects of our fermionic DM $N_1$. In vanilla Scotogenic model, the elastic scattering of the DM $N_1$ with nucleon happens via the one-loop diagram as shown in the left panel of Fig.~\ref{fig:DD-diagram}. This loop diagram is suppressed due to the smallness of Yukawa coupling $Y^\nu$. In the dynamical Scotogenic model, in addition to this loop-level process there is a tree level process for DM elastic scattering on nucleons through $t$-channel mediation of Higgs bosons $h,H$. The spin independent cross section, $\sigma_{SI}$, for process $N_{1} X \to N_{1} X$ where X is nucleon is given by,
\begin{align}
\sigma_{SI}=\frac{4}{\pi}\mu_{r}|A^{SI}|^{2}
\label{eq:DD},
\end{align}
where $\mu_{r}=\frac{M_{N_{1}}M_{X}}{M_{X}+M_{N_{1}}}$($M_{X}$ is the nucleon mass) and $A^{SI}$ is nucleon amplitude which depends on the DM interaction with nucleons and the nuclear form factors. To evaluate nucleon amplitude $A^{SI}$, we use MicroOmegas \cite{Belanger:2014vza}. In Fig.~\ref{fig:DD}, we show the spin-independent direct detection cross-section versus the DM mass $M_{N_1}$. Clearly there are solutions with the correct DM relic and direct detection cross-sections. The green and red band in Fig.~\ref{fig:DD} represents the constraint from XENON-1T and LUX-Zeplin \cite{XENON:2018voc,LZ:2022ufs} experiment. All points are allowed by neutrino mass constraint, astrophysical constraint, lepton flavor violation and the relic density for these points vary in between $\Omega_{N_1} h^2 \in 0.0001-0.1226$. In Fig.~\ref{fig:DD}, we can see that larger spin independent cross section gets severly constraint from both XENON-1T and LUX-Zeplin experiment.

As we have emphasized before for our chosen parameter space, the DM annhilation to majoron plays an important role to produce the right relic abundance. As the majoron being singlet and massless, it neither undergoes decay to SM particles nor emits photons in the final state. Therefore, we do not have any constraints on the DM mass from DM indirect detection experiments.

\section{Conclusion}
\label{sec:conclusion}
In Scotogenic models, the lightest right-handed neutrino $N_1$ is a thermal DM candidate. In the vanilla Scotogenic scenario, there is a strong correlation between  the LFV process $\mu \to e\gamma$ and the relic abundance of $N_1$ governed by the $N_1 N_1 \to \ell^\pm\ell^\mp/\nu_{l}\nu_{l}$ process. Taking into account the neutrino mass constraint and setting the neutrino oscillation parameters to the best fit-values, we observe that the LFV constraint forces the DM to be produced in over-abundance in a majority of parameter space. In the dynamical Scotogenic model, other processes involving the massless majoron mode $J$, as well as the standard and non-standard Higgs bosons $h,H$ in the final state also  substantially contribute to the DM relic abundance. We find that additional processes such as $N_1 N_1 \to J J, J\ h/H$ are dominant enabling the DM to be in thermal equilibrium with thermal bath until a  later time as compared to the $N_1 N_1 \to \ell^\pm\ell^\mp/\nu_{l}\nu_{l}$ process. This reduces the DM over-abundance and hence a correct relic density can be achieved.  In addition to $\mu \to e \gamma$, process such as $\mu \to e J$ can give strong constraint on the viable parameter space. We examined the viable parameter space of the model for the DM mass in the 100 \rm{GeV}-10 \rm{TeV} range satisfying the neutrino data, LFV bounds on $\mu \to eJ$ and $\mu \to e \gamma$, as well as astrophysical and cosmological constraints including stellar cooling and $N_{\rm eff}$. We have also found substantial points in parameter space satisfying the observed DM relic density and also in agreement with direct detection searches. The DM indirect detection does not constrain our chosen parameter space as the DM dominantly annihilates to massless states majoron.

\begin{acknowledgements}
MM acknowledges the support from the Indo-French Centre for the Promotion of
Advanced Research (Grant no: 6304-2).The work of S.M. is supported by KIAS Individual Grants (PG086001) at Korea Institute for Advanced Study. AR acknowledges SAMKHYA: High-Performance Computing Facility provided by the Institute of Physics (IoP), Bhubaneswar.
\end{acknowledgements}
\bibliographystyle{utphys}
\bibliography{bibitem}
\end{document}